\documentclass[]{aa}
\usepackage{graphicx}
\usepackage{psfig}
\usepackage{lscape}
\usepackage{rotating}
\usepackage{longtable}
\usepackage{txfonts}
\input epsf

\begin{document}
   \title{Reduced Wolf-Rayet Line Luminosities at Low 
Metallicity\thanks{Based on observations made with ESO  
Telescopes at the La Silla Observatory  under programme ID 074.D-0696(A)}}

\titlerunning{Low WR line luminosities at low metallicity}

   \author{Paul A. Crowther  \and           L.J. Hadfield}

   \offprints{Paul Crowther (Paul.Crowther@sheffield.ac.uk)}

         \institute{Department of Physics \& Astronomy, University of Sheffield, 
Hicks Building, Hounsfield Rd,
   Sheffield, S3 7RH, United Kingdom
             }

\authorrunning{Crowther \& Hadfield}

   \date{Received; accepted}

   \abstract{New NTT/EMMI spectrophotometry of single WN2--5 stars in
the Small and Large Magellanic Clouds are presented, from which He\,{\sc
ii} $\lambda$4686 line luminosities have been derived, and compared
with observations of other Magellanic Cloud Wolf-Rayet 
stars from the literature. 
SMC WN3--4  stars possess line luminosities which are a
factor of 4 times lower than LMC counterparts, incorporating 
several binary SMC WN3--4 stars from the literature. Similar results
are found for WN5--6 stars, despite reduced statistics, 
incorporating observations of single LMC WN5--9 stars 
from the literature.  C\,{\sc iv} $\lambda$5808 line luminosities of
carbon sequence WR stars in the SMC and IC\,1613 (both WO subtypes) 
from the recent literature are  a factor of 3 lower than LMC WC stars from
Mt Stromlo/DBS spectrophotometry, although similar
results are also obtained for the sole LMC WO star. We demonstrate how 
reduced  line luminosities at low metallicity follow  naturally 
if WR winds are metallicity-dependent, as recent empirical and 
theoretical results suggest.  We apply mass loss-metallicity 
scalings to atmospheric non-LTE models of Milky Way  and LMC WR stars 
to predict
 the wind signatures of WR stars in the metal-poor star forming WR galaxy
I\,Zw~18. WN He\,{\sc ii} $\lambda$4686 line luminosities are 
7--20 times lower 
than in metal-rich counterparts of identical bolometric luminosity, whilst
WC C\,{\sc iv} $\lambda5808$ line luminosities are 3--6 times lower.
Significant He$^{+}$ Lyman continuum fluxes are
predicted for metal-poor early-type WR stars. Consequently, our results
suggest the need for larger
population of WR stars  in I\,Zw~18 than is presently assumed, particularly
for WN stars,  potentially posing a severe
challenge to evolutionary models at very low metallicity. Finally, reduced 
wind strengths from WR stars at low metallicities impacts upon the immediate 
circumstellar  environment of long duration GRB afterglows, particularly 
since the host galaxies of high-redshift GRBs tend to be metal-poor.
\keywords{stars: Wolf-Rayet -- galaxies: stellar content -- galaxies: 
individual: I\,Zw~18}

}

   \maketitle
%

\section{Introduction}

Wolf-Rayet stars -- subdivided into nitrogen (WN) and carbon (WC) sequences
-- are the evolved descendants of the O stars, and possess
strong, broad, emission lines due to their dense, 
stellar winds. WR galaxies
represent a subset of emission-line galaxies with active massive star
formation via the direct signature of WR stars.  To date, over a hundred WR
galaxies are known in the nearby universe (Schaerer et al. 1999) via a
broad C\,{\sc iii} $\lambda$4650/He\,{\sc ii} $\lambda$4686
(blue bump) and/or C\,{\sc iv} $\lambda$5808     (yellow bump) emission
features seen in the integrated optical spectrum of individual sources.
Indeed, broad He\,{\sc ii} $\lambda$1640 emission, attributed to WR stars, 
can be easily seen in the
average rest-frame spectrum of $z\sim 3$ Lyman Break Galaxies (LBGs, Shapley et al.
2003).

The O star content of WR galaxies is typically derived indirectly using
nebular hydrogen emission line fluxes to determine the total number of
ionizing photons, from which the equivalent number of O7V stars is commonly
calculated. The actual O star content depends primarily upon the age and
mass function, and relates to the equivalent number of O7V stars via a
correction factor (Vacca 1994). As such, this quantity is fairly
metallicity {\it independent}, although O7V stars at lower metallicity 
will likely have higher temperatures -- and so higher Lyman continuum 
fluxes -- than their high metallicity counterparts (Massey et al. 2005; 
Mokiem et  al. 2006).

In contrast, the WR content is routinely obtained merely by dividing the
observed emission bump fluxes, corrected for reddening and distance, by
average line luminosities of Milky Way and Large Magellanic Cloud WR stars
(Schaerer \& Vacca 1998). As such, line luminosities of WR stars are {\it
assumed} to be metallicity independent.  Following this technique, 
the WR content of metal-rich (e.g. Mrk 309: Schaerer et al. 2000)  and 
metal-poor (e.g. I\,Zw~18: Izotov et al. 1997) galaxies have been derived 
and compared, generally successfully, with evolutionary synthesis models.

Recent observational and theoretical evidence suggests WR winds depend upon
the heavy metal content of the parent galaxy (Crowther et al. 2002;
Gr\"{a}fener \& Hamann 2005; Vink \& de Koter 2005).  
Since WR stars are believed to be
the immediate precursors of some long-soft Gamma Ray Bursts (GRBs) the
circumstellar environment of low metallicity GRBs is expected to differ
substantially from those in metal-rich regions (Eldridge et al. 2005).
If WR winds depend upon metallicity, do their line luminosities?  

In the present study, we present
new optical spectrophotometry of single, early-type WN stars in the Large
and Small Magellanic Clouds in Sect~\ref{obs}. The Magellanic Clouds
were selected on the basis of their known distances, resolved stellar content
and low interstellar reddenings.
Indeed, reduced He\,{\sc ii}
$\lambda$4686 line luminosities for SMC WN stars ($\log$~O/H$+12\sim$8.1)  
relative to the LMC ($\log$~O/H$+12\sim$8.4) are obtained in
Sect.~\ref{line}, incorporating observations of late-type WN stars from the
recent literature.  We also demonstrate that carbon sequence 
WR stars at low metallicity -- dominated by the rare WO subclass -- also possess lower 
line luminosities than typical LMC counterparts based upon 
spectrophotometry of single and binary WC stars, published 
in part by Crowther et al. (2002). In Sect~\ref{models}, we
consider whether reduced WR line luminosities are expected at lower
metallicities for metallicity dependent WR winds. The results of our study
are discussed in Sect.~\ref{discussion}, with particular application to
I\,Zw~18, and brief conclusions are drawn in Sect~\ref{summary}.

\section{Observations of Magellanic Cloud WN stars}\label{obs}

We present new spectrophotometry of single LMC and SMC early-type WN 
stars (Foellmi et al. 2003a, b). 
Photometric properties of 42 LMC and 7 SMC WN stars
are presented in Table A1 in the Appendix. 
LMC WR catalogue numbers
include both Breysacher (1981, Br) and Breysacher et al. (1999, BAT99), for
completeness, whilst SMC WR catalogue numbers follow Massey et al. (2003).
The use of single stars at known distances permits a uniform method 
of deriving interstellar reddenings and line luminosities. In 
addition, average `generic' spectra may be obtained,
of potential application in synthesising the bumps in Wolf-Rayet galaxies.

\subsection{Observations and data reduction}
\label{Spectrophotometry}

We have observed our LMC and SMC sample using the ESO Multi Mode Instrument
(EMMI) on the 3.5-m New Technology Telescope (NTT), La Silla, Chile.  The
detector consists of two 2048$\times$4096 MIT/LL CCDs which were binned by
a factor of 2 in both the spatial and dispersion direction.  Data was
acquired using the RILD (Red Imaging and Low Dispersion 
Spectroscopy) mode of EMMI with grism \#3 (2.8\AA\,pixel$^{-1}$ 
dispersion) resulting in a
wavelength coverage of 3800--9070\AA\, with a resolution of 9.3\AA, as
measured from comparison arc lines.

Conditions during data acquisition were photometric and seeing estimates
varied between 0.6 and 1.1\arcsec.  These good conditions together with a
wide, 5\arcsec\ slit suggests that complete transmission was achieved, and
that the data are truly spectrophotometric.  On-source exposure times ranged
from 60s for the brightest objects (e.g. BAT99-117) to 1260s for the faintest
(e.g. BAT99-23).

The data were reduced following the usual procedure i.e. bias subtracted,
extracted and flux / wavelength calibrated using packages within {\sc iraf}
and {\sc starlink}, except that flat fielding was not carried out due to
spurious structure in all calibration frames.
Care was taken during the extraction process to ensure the entire
profile was extracted.

Absolute flux calibration was achieved by observing the spectrophotometric
standard stars GD\,50, GD\,108, G\,158-100 and Feige~110. A comparison
between several standards taken at regular intervals during each night
suggests that flux calibration of the data is accurate to $\pm$5\%.

WN spectral types follow Foellmi et al., except for BAT99-50 and 
BAT99-73,  
which 
are revised from WN4 to WN5 on the basis that N\,{\sc iv} $\lambda$4058 
$\gg$ N\,{\sc v} $\lambda$4610/N\,{\sc iii} 
$\lambda$4640.  

During the reduction of SMC--WR11 it became apparent that its spectrum was
contaminated by a red line-of-sight companion $\sim$1.2 arcsec to the 
west, which appears slightly brighter than  the WR star in R-band 
acquisition  images. We have extracted two apertures for this target: (i) 
the first  provides spectrophotometry
for the composite spectrum, but will overestimate the derived reddening
because of the red companion; (ii) the second attempts to 
isolate the WN star, whose colours will provide a more robust reddening. 
Both entries are given for SMC-WR11 in Table~A1.  The presence of
a red companion naturally explains the unusually bright 2MASS JHK 
photometry for this star relative to other SMC WN stars.


\subsection{Synthetic-Filter photometry}

We have derived synthetic magnitudes for our LMC and SMC sample by
convolving the calibrated spectra with {\it synthetic} Gaussian filters.  
The filters have central wavelengths of $\lambda_b = 4270 \AA, \lambda_v =
5160 \AA\ \mbox{and} \, \lambda_r = 6000 \AA$ and mimic the response of
narrow-band {\it b, v} and {\it r} filters (Smith 1968;  Massey 1984),
 with zero-points adopted that reproduce the synthetic magnitudes for
the dataset of  Torres-Dodgen \& Massey (1988). In contrast with Schmutz 
\& Vacca (1991), the intrinsic colours introduced in Sect.~\ref{reddening} 
include spectral lines, so we consider synthetic rather than
monochromatic (line-free) magnitudes.
An accuracy in the flux calibration of $\pm 5$\% translates to a photometric
uncertainty of $\pm 0.05$~mag. As discussed above, synthetic magnitudes for
SMC-WR11 are less reliable than other targets, due to its near neighbour.

Synthetic magnitudes and colours 
are presented in Table A1, together with differences between this
study and that published by Torres-Dodgen \& Massey (1988).  For the 31 LMC
stars in common, we find satisfactory agreement between
synthetic magnitudes and colours, with average differences of $\Delta v
=v_{\rm EMMI}- v_{\rm TM88} = -0.13\pm0.21$ mag and $\Delta(b-v) =
(b-v)_{\rm EMMI} - (b-v)_{\rm TM88} = -0.02\pm0.06$ mag, respectively.  In a
minority of cases differences were quite significant, with $\Delta v$ for
BAT99-37 and BAT99-128 differing by --0.86 and +0.76 mag, respectively. 
For these, Torres-Dodgen \& Massey (1988)  note that their results are
uncertain due to non-photometric conditions. For the SMC sample,
Torres-Dodgen \& Massey only provide $u$ and $b$ magnitudes. For three WN
stars in common, $\Delta b = b_{\rm EMMI} - b_{\rm TM88}$ = +0.12$\pm$0.04
mag.

\begin{figure}[h]
\centerline{\psfig{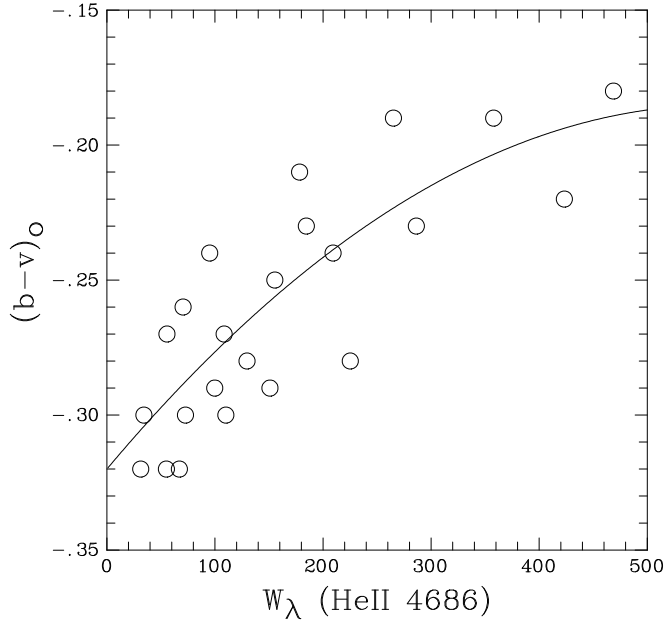}}
\caption{Comparison between theoretical
He\,{\sc ii} $\lambda$4686 equivalent widths and $(b-v)_{0}$ colours
from the Smith et al. (2002) grid of WN models of Solar, LMC and SMC 
metallicities with 40kK $\leq T_{\ast} \leq$ 100kK 
(open symbols), together with a quadratic fit  (solid line).} 
\label{colour} 
\end{figure}

\subsection{Interstellar reddening}\label{reddening}

Estimates of the interstellar reddening towards our sample have been
derived from comparison between the observed colours and theoretical
non-LTE model predictions, following Schmutz \& Vacca (1991). Ideally, one
would use $u - b$, but since our data did not extend
far enough into the blue, we have instead based our reddening estimates on
the standard $b - v$ colour index.

It is well known that non-LTE WR models with stronger emission lines
display flatter optical continuum spectral energy distributions
(Schmutz \& Vacca 1991). Using the grid of WN non-LTE model spectra
generated by Smith et al. (2002) for evolutionary synthesis calculations,
we have derived the following approximate relationship between $(b -
v)_0$ and the He\,{\sc ii} $\lambda$4686 equivalent width, $W_{\lambda}$
(in \AA)
\[ 
\begin{array}{rll} 
(b - v)_0 &= -0.32 + 0.000476 \, W_{\lambda}
- 4.2 \times 10^{-7} (W_{\lambda})^{2}&{\rm mag}\\ 
\end{array} 
\] 
which is
presented in Fig.~\ref{colour}, together with individual data points from
the 40kK $\leq T_{\ast} \leq$ 100kK models for Solar, LMC and SMC
metallicities. $E_{B-V}$  followed from the standard relation
$E_{B-V}/E_{b-v} = 1.21$ (Lundstr\"{o}m \& Stenholm 1984).

The scatter in the correlation between $(b - v)_0$ and $W_{\lambda}$(4686)
suggest that $(b - v)_0$ is accurate to $\sim 0.04$~mag.  Combining this
with the estimated uncertainty of $\pm 0.05$~mag in the observed {\it b} and
{\it v} colours, we expect that the typical uncertainty in derived
$E_{B-V}$ values will be $\sim\pm 0.06$mag.

Derived $E_{B-V}$ values for LMC and SMC WNEs are presented in 
Table~A1.  We have compared reddening estimates for LMC WN stars to those
published by Schmutz \& Vacca (1991).  For the 28 objects in common with
both studies we find reasonable agreement, $\Delta E_{B - V} = E_{B-V}({\rm
this~study}) - E_{B-V}({\rm SV91}) = 0.07\pm0.08$~mag on average.
 For the SMC WN stars, Massey \& Duffy
(2001) and Massey et al. (2003) assumed $(B-V)_{0} = -0.32$~mag. 
Excluding SMC-WR11, which is contaminated by a red companion, we obtain
$\Delta E_{B - V} = E_{B-V}({\rm this~study}) - E_{B-V}({\rm MD01}) = 
-0.05\pm0.04$~mag.

With interstellar reddenings determined, we then fit
Gaussian line profiles to individual  He\,{\sc ii} $\lambda$4686 emission lines
using {\sc dipso} (Howarth et al. 2003),
revealing line fluxes, FWHM and equivalent widths, also presented in 
Table A1. We assume a distance modulus
of $\mu_{\rm LMC}$ = 18.45$\pm$0.06
mag (49\,kpc), which represents the mean of 7 techniques discussed by 
Gibson (2000), and $\Delta \mu = 
\mu_{\rm SMC} - \mu_{\rm LMC}$ = 0.51$\pm$0.03 mag, based upon three 
standard candle techniques (Udalski et al. 1999), implying $\mu_{\rm SMC}$ 
= 18.96 $\pm$0.07 mag (62\,kpc), in reasonable agreement with the recent 
eclipsing binary determination by  Hilditch et al. (2005) of $\mu_{\rm 
SMC}$ = 18.91 $\pm$0.03 mag. Line luminosities then follow from our derived 
reddenings and our assumed distances.

\begin{table}[h]
\caption{Mean optical line luminosities ($\pm$ standard deviations)
of single Magellanic Cloud WN stars (units 
of 10$^{34}$  erg/s). 
For comparison, Schaerer \& Vacca 
(1998) obtained 5.2$\pm2.7 \times 10^{35}$ erg/s and 1.6$\pm 1.5\times 
10^{36}$ erg/s for He\,{\sc ii} $\lambda$4686 in
Galactic/LMC WN2--4 and  WN6--9 stars, respectively. 
Values in parenthesis
additionally include three SMC WN3--4 binaries.}\label{average_wn}
\begin{tabular}{
l@{\hspace{3mm}}
l@{\hspace{2mm}}
l@{\hspace{3mm}}
l@{\hspace{2mm}}
l@{\hspace{3mm}}
l}
\hline
Line & \multicolumn{2}{c}{WN2--4} & \multicolumn{2}{c}{WN5--6} & WN7--9 \\
                                      & LMC & SMC & LMC & SMC & LMC \\
Number     & 36 & 5 (8) & 15 & 2 & 9 \\
\hline
N\,{\sc iv} $\lambda$4058             &  2$\pm$2&             & 26$\pm$24   & 10$\pm$9 & 17$\pm$9 \\
N\,{\sc v} $\lambda\lambda$4603--20   & 16$\pm$20 & 1.4$\pm$0.9 &             &           &           \\
                                      &            & (2.4$\pm$2.1) & & & \\
N\,{\sc iii} $\lambda\lambda$4634--41 &           &             & 25$\pm$27   & 4$\pm$4   & 38$\pm$19 \\
He\,{\sc ii} $\lambda$4686            &  84$\pm$71 & 5.8$\pm$3.4& 175$\pm$166 & 43$\pm$49 & 72$\pm$67 \\
                                      &                 & (17$\pm$16) & & & \\
\hline
\end{tabular}
\end{table}

\section{Reduced line luminosities of WR stars at low metallicity}\label{line}

\subsection{He\,{\sc ii} $\lambda$4686 in WN stars}\label{4686}

It is well known that SMC WN stars possess relatively weak, narrow
emission lines with respect to WN stars in the LMC and Milky Way (Conti et
al. 1989). We now investigate whether their line luminosities also differ
between these galaxies.

We have combined our NTT/EMMI datasets with mid to late-type LMC WN stars from Crowther \&
Smith (1997) uniformly degraded in spectral  resolution to $\sim$10\AA\ to mimic the  EMMI datasets, 
supplemented by  HST/FOS observations of R136a1, a2 and a3 (see Crowther \& Dessart 1998).

\begin{figure} 
\centerline{\psfig{figure=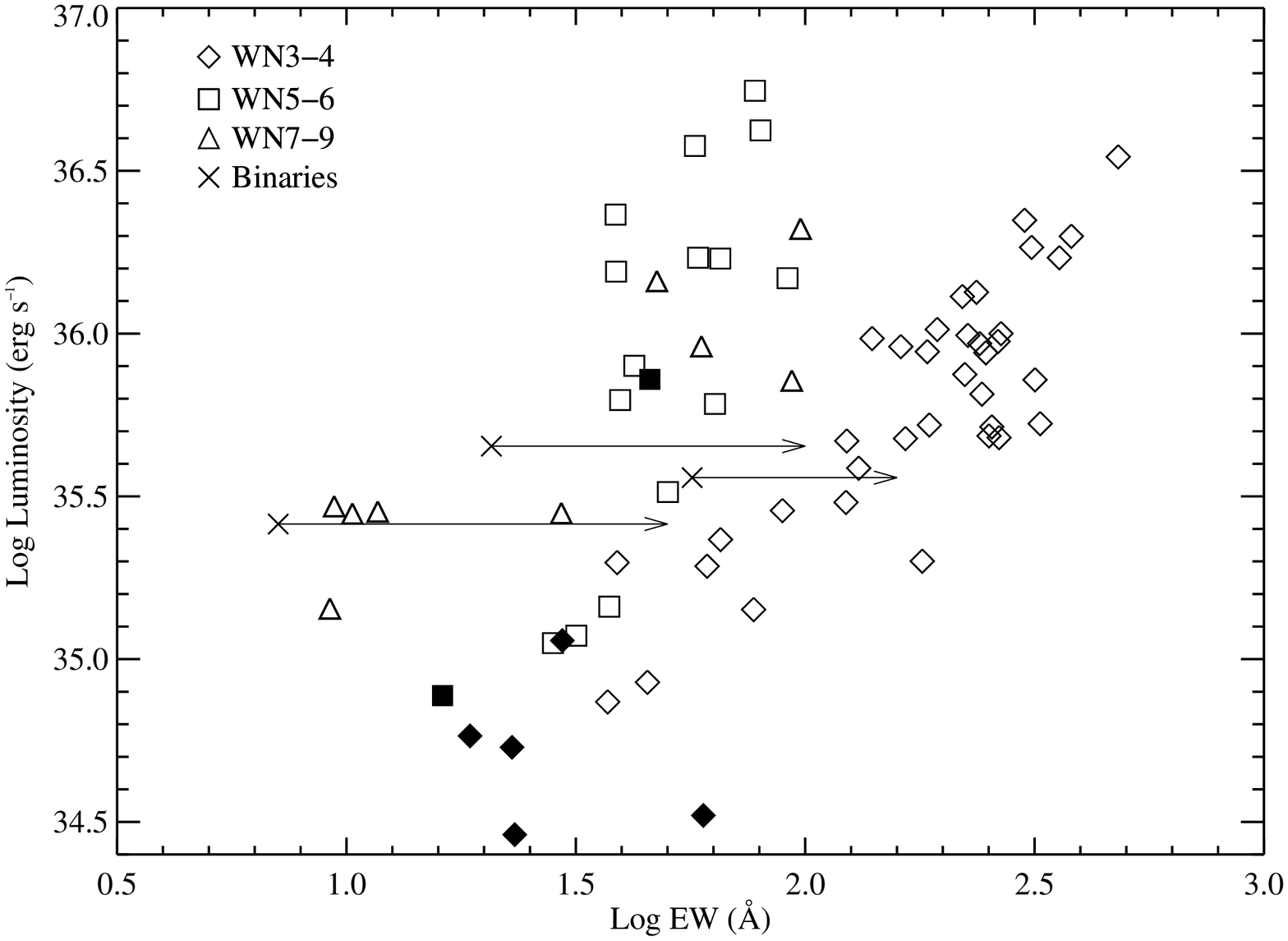,width=8.8cm,angle=90}}
\centerline{\psfig{figure=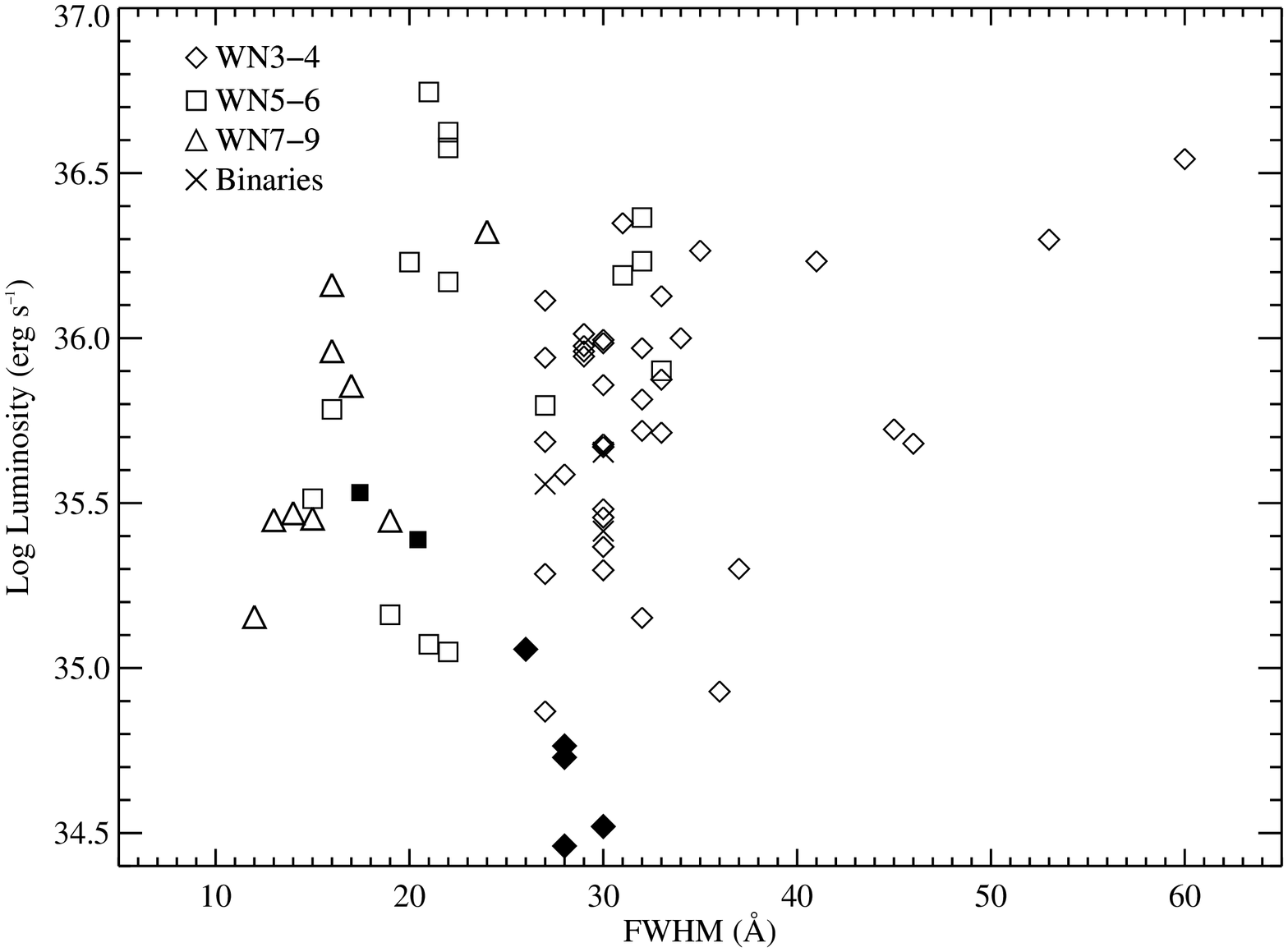,width=8.8cm,angle=90.}} 
\caption{Comparison between the He\,{\sc ii} $\lambda$4686 line
luminosities of single LMC (open) and SMC (filled) early, mid and 
late-type WN stars versus equivalent width (top) and FWHM (bottom),
supplemented by three SMC WN3--4 binaries (crosses). Data 
are taken from our NTT/EMMI datasets, supplemented by LMC late-type WN 
stars from Crowther \& Smith (1997) and Crowther \& Dessart (1998), 
uniformly degraded to the spectral resolution of EMMI (10\AA), 
explaining the deficiency at small FWHM. Data for the three SMC WN3--4 
binaries are from Torres-Dodgen \& Massey (1998), for 
which an estimate of the line dilution has been made on the basis of  
$M_{v}=-4$ mag for WR3 and $M_{v}=-$5 mag for WR6 and
WR7 (Foellmi et al 2003a), although we obtain $M_{\rm}$=--3.9$\pm$0.7 mag 
for our single SMC WN3--4 stars. } \label{wn} \end{figure}

Results for individual stars are presented in Fig.~\ref{wn}, where
we compare He\,{\sc ii}  $\lambda$4686 line luminosities with equivalent
widths (top) and FWHM (bottom).   Fig.~\ref{wn} supports the conclusions of 
Conti et al. (1989) that  SMC stars have weaker and narrower He\,{\sc ii} 
emission lines than their LMC counterparts  for each WN subclass, 
although we have further restricted our sample to apparently single WN stars. 
This was primarily motivated to ensure a uniform
method of deriving reddenings and line luminosities. 
Nevertheless, since our SMC sample comprises only seven WN stars in total, 
we have assessed the impact of  including SMC WN binaries in our study. 
There are three additional WN3-4+O systems (SMC WR3, WR6, WR7) for 
which we have used the Torres-Dodgen
\& Massey (1988) spectrophotometry, plus reddening estimates from
Crowther (2000).
Their inclusion leads to a significant increase in the SMC WN2--4 average 
line luminosity. Nevertheless, it is clear that He\,{\sc ii} $\lambda$4686 
line luminosities of  SMC stars are typically factors of  4--5 lower than 
their LMC counterparts for each WN subclass.

As previously discussed by Schaerer \& Vacca (1998)
there is a substantial scatter in the observed He\,{\sc ii} 
$\lambda$4686 line luminosities of
LMC WN stars.   FWHM( He\,{\sc ii} $\lambda$4686) clusters around 
$\sim$28\AA\ for 
early WN stars in the SMC, whilst their LMC counterparts span a much 
broader range 
from 27--60\AA, although their mean (32\AA) is not so different from
those in the SMC. Mean line luminosities for early (WN2--4), mid (WN5--6) 
and late (WN7--9) subtypes for each galaxy 
are presented in Table~\ref{average_wn}, 
where quoted uncertainties reflect the standard deviation on  individual 
measurements,  with values in parenthesis for early SMC WN stars 
additionally including binaries.

Our mean LMC WN2--4 He\,{\sc ii} 
$\lambda$4686 line luminosity is  $\sim$60\% higher than compiled by Schaerer \& 
Vacca (1998), which was based on a reduced sample of LMC and Galactic 
stars, using inhomogeneous reddening and distance determinations. Our
mean LMC WN5--6 and WN7--9 He\,{\sc ii} $\lambda$4686 line luminosities
straddle that determined by Schaerer \& Vacca (1998) for Galactic/LMC 
WN6--9 stars (WN5 stars were excluded from their sample). Chandar et al.
(2004) quote similar results using LMC results from Conti \& Morris (1990).
Vacca (1992) determined an average late-type WN line luminosity of 1.7$\times 
10^{36}$ 
erg/s that is in good agreement with our WN5--6 mean, and represents a
good match to that resulting for the 30 Doradus (LMC: Vacca 1992)
and NGC~604 (M33: Terlevich et al. 1996)  complexes from direct WN star 
counts. 

\begin{figure}[h]
 \centerline{\psfig{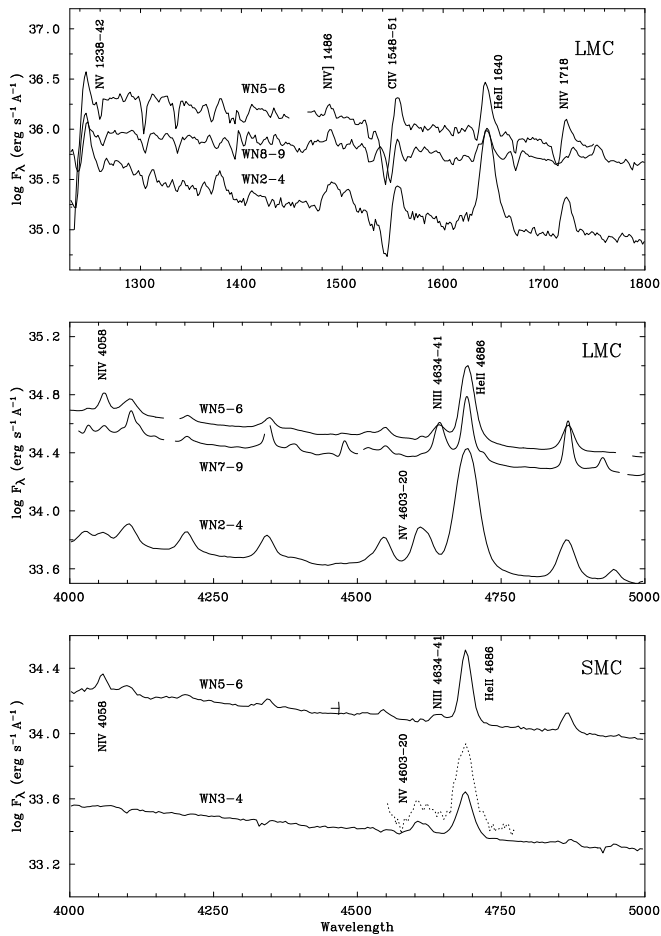}} 
\caption{`Generic' UV and optical spectra of single, 
early (WN2--4), mid 
(WN5--6), and late (WN7--9) nitrogen sequence spectra for the LMC  
and SMC. The UV late WN spectrum is representative of WN8--9 stars, 
since no single WN7 stars have been observed with IUE/SWP.  An
additional  spectrum for SMC WN3--4 stars (dotted lines) 
incorporates the binaries SMC-WR3, WR6 and WR7, 
for which an attempt has been made to 
correct for the continuum of the O star companion (see text).
}\label{flux} 
\end{figure}

\begin{figure}[h]
\centerline{\psfig{figure=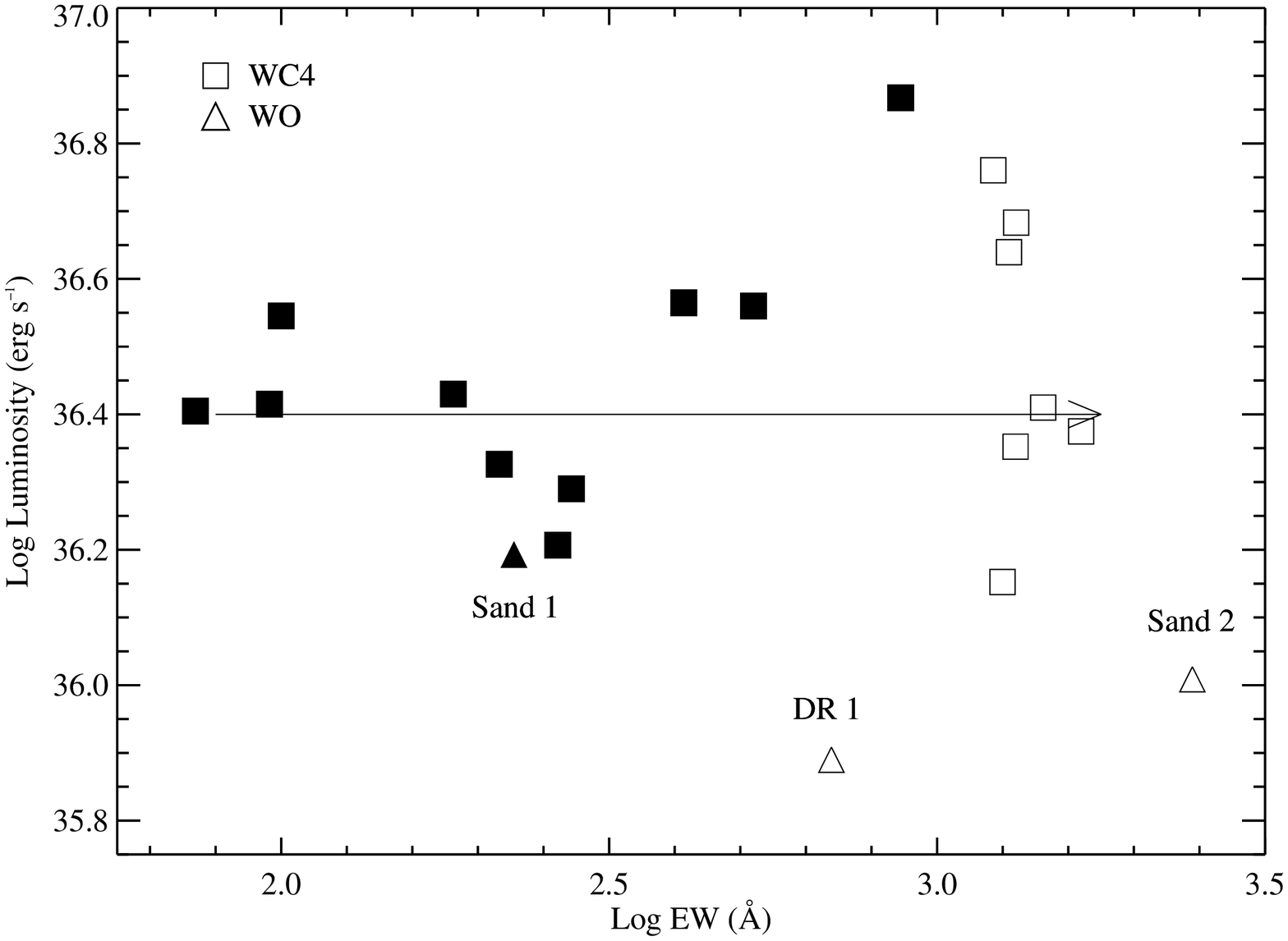,width=8.8cm,angle=90}}
\centerline{\psfig{figure=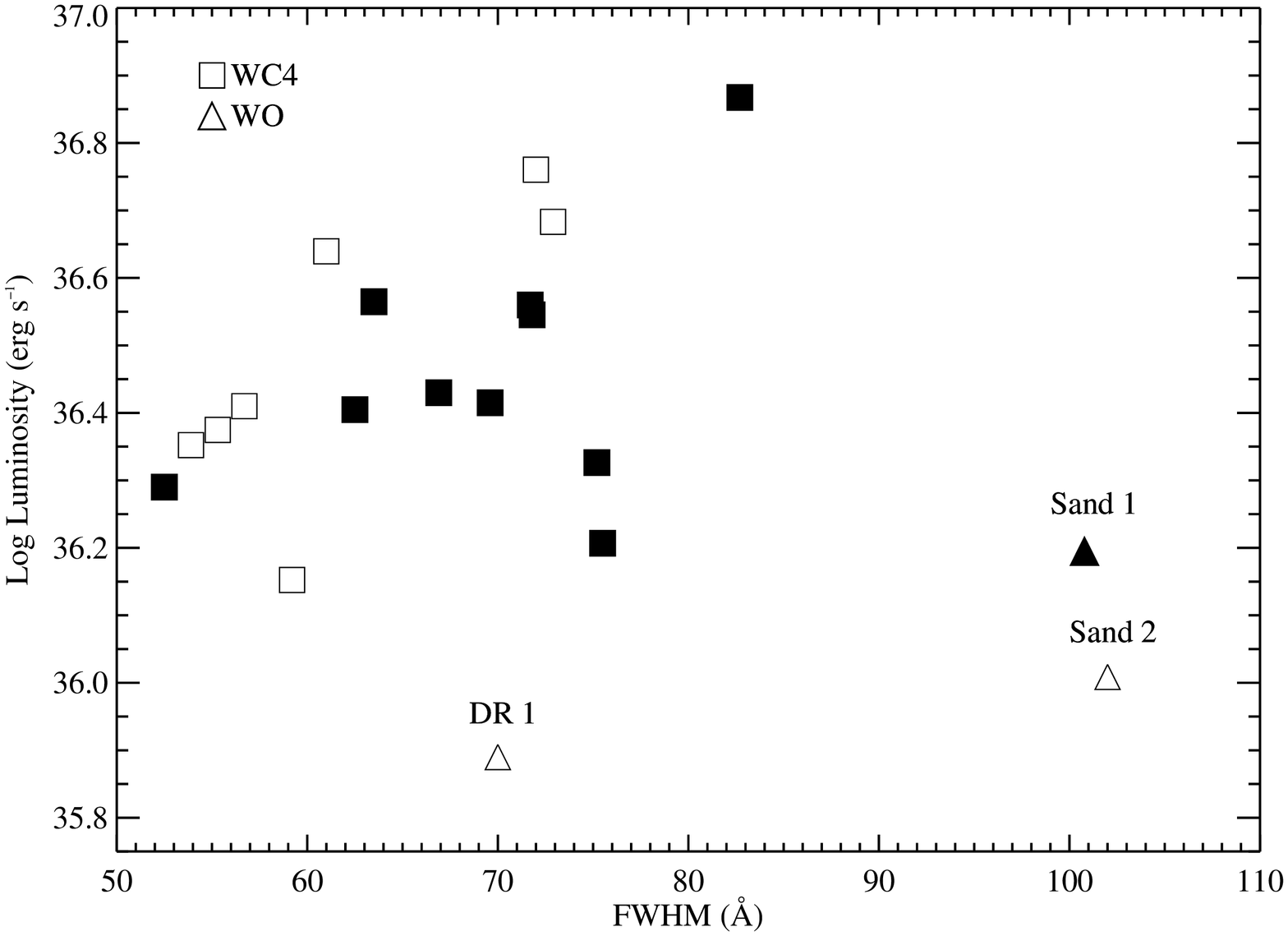,width=8.8cm,angle=90.}}
\caption{Comparison between C\,{\sc iv} $\lambda$5808 line 
luminosities  of single (open) and binary (filled) WC4 (circles) and WO 
(triangles) stars in the LMC, SMC (Sand 1) and IC~1613 (DR1). WO 
observations  are taken from  Kingsburgh et al.  (1995) and Kingsburgh \& 
Barlow (1995), WC4 observations are from Mt Stromlo 2.3m/DBS (see text). 
An estimate of  line dilution is indicated for BAT99-10 (WC4+O) for which 
Walborn et al. (1999) have obtained spatially resolved HST/FOS blue 
spectroscopy of the WC4 component (their HD~32228-2). 
}\label{wc}
\end{figure}


We have explicitly excluded HD~5980 (SMC WR5) from Fig.~\ref{wn} and 
Table~\ref{average_wn}
since this 19.3\,day WN+OB binary underwent an LBV eruption during 1994, with
a B1.5Ia$^{+}$ or WN11 spectral type during outburst (Barba et al. 1995).
Indeed, the He\,{\sc ii} $\lambda$4686 line in HD~5980 is highly phase-dependent as a
result of strong excess emission arising from its wind-wind collision zone (Moffat
et al. 1998). 

Nevertheless, for completeness we have estimated its $\lambda$4686
line luminosity from spectrophotometry of Torres-Dodgen \& Massey (1988) plus our
own AAT (RGO spectrograph) and Mt Stromlo 2.3m (DBS spectrograph) observations from
circa 1981-84 to 1997. Adopting an interstellar reddening of E(B-V)=0.07 (Crowther 2000)
we have estimated $\lambda$4686 line luminosities of 5.5--15$\times 10^{36}$ erg/s,
unmatched during outburst by any Magellanic Cloud WN star.
Koenigsberger et al. (2000) present UV 
spectroscopic observations between 1979 and 1999,  also revealing 
a factor of $\sim$3 increase in  He\,{\sc ii} $\lambda$1640 flux over that interval.


  In addition, since our primary sample is drawn exclusively from 
single stars we have constructed `generic' early, mid and late WN optical 
spectra for each galaxy. Individual spectra were dereddened
allowing for a uniform foreground Galactic (Seaton et al. 1979)
extinction of E(B-V)=0.05 mag, with the remainder according to either LMC
(Howarth 1983) or SMC (Bouchet et al. 1985) extinction laws.
The co-added average spectrum was subsequently scaled to the mean 
He\,{\sc ii} $\lambda$4686 line luminosity, in each case. These are presented in 
Fig.~\ref{flux},  and reinforce the reduced line luminosities of SMC 
stars,
relative to LMC counterparts for each WN subclass. For the SMC WN3--4
subtypes, the dotted spectrum in the region of He\,{\sc ii} $\lambda$4686
additionally includes the three binaries SMC WR3, WR6 and WR7 in which we have
attempted to take account of their O star continua by assuming Kurucz (1993)
models adjusted such that 
$M_{v}=-4$ mag for WR3 and $M_{v}=-$5 mag for WR6 and
WR7  following fig.~16 of Foellmi et al (2003a).\footnote{
Based upon our spectrophotometry
and E(B-V) values, we obtain $M_{\rm}$=--3.9$\pm$0.7 mag for single SMC 
WN3--4 stars.}
The generic spectra have 
potential use in synthesising the WR bumps in low metallicity 
star forming galaxies (e.g. Hadfield \& Crowther, in preparation).

\subsection{He\,{\sc ii} $\lambda$1640 in WN stars}\label{1640}

Since He\,{\sc ii} $\lambda$1640 from WR stars is seen both in nearby
starburst clusters (Chandar et al. 2004) and the average LBG  spectrum 
(Shapley et al. 2003), what is the He\,{\sc ii} 
$\lambda$1640/$\lambda$4686 flux ratio of WN stars? 
Conti \& Morris (1990) compared F(He\,{\sc ii} $\lambda$1640)/F(He\,{\sc ii}
$\lambda$4686)
for a sample of LMC WN stars, revealing $\sim$7.6 (see also Chandar et al.
2004). Schaerer \& Vacca (1998) obtain 7.95$\pm$2.47  and 7.55$\pm$3.52 for early and 
late WN stars in the Milky Way and LMC.

29 LMC stars from
our sample were observed with IUE/SWP using the large aperture (LAP)
at low resolution (LORES), from which we have obtained
an average ratio of F(He\,{\sc ii} $\lambda$1640)/F(He\,{\sc ii} $\lambda$4686) of 
9.9$\pm$3.2\footnote{There is a marginal trend to a lower 
$\lambda$1640/$\lambda$4686 ratio from early-WN's (10.8, $\sigma$=3.2, 
N=18) through 
mid-WN's (8.4, $\sigma$=3.5, N=7) to late-WN's (8.4, $\sigma=$1.5, N=4)}. 
For 3 single SMC
stars with low resolution, large aperture IUE SWP spectrophotometry, we 
obtain a similar average ratio of 11.9$\pm$1.0.
Additionally, we have compared the theoretical F(He\,{\sc ii} 
$\lambda$1640)/F(He\,{\sc ii} $\lambda$4686) ratio from the non-LTE model WN grid
of Smith et al. (2002) discussed above, revealing 9--10 for SMC to 
Milky Way  metallicities ($W_{\lambda} 1640/4686 \sim$ 0.32). 
Consequently, we suggest F(He\,{\sc ii} 
$\lambda$1640)/F(He\,{\sc ii} $\lambda$4686) = 10 $\pm$1 for
WN stars with metallicities between 1/5$Z_{\odot}$ and 1$Z_{\odot}$.

 Generic  ultraviolet early, mid, and
late WN spectra for LMC metallicities are also included in 
the top panel of Fig.~\ref{flux}. These were obtained from IUE/SWP 
datasets, supplemented by HST/GHRS observations of R136a1, a2 and a3 for 
WN5--6 subtypes (Heap et al. 1994; Crowther \& Dessart 1998), degraded 
to the low resolution IUE observations. Unfortunately,  no single LMC WN7 
stars have been observed with IUE/SWP, so our late-type WN average is drawn 
from solely WN8--9 stars. Since only a subset of our LMC WN sample possess
ultraviolet spectroscopy, these have been adjusted relative to the 
generic optical WN spectra, such that F(He\,{\sc ii} $\lambda$1640)/F(He\,{\sc 
ii}  $\lambda$4686) = 10.

\subsection{C\,{\sc iv} $\lambda$5808 in WC stars}\label{5801}

Do low metallicity WC  stars also possess reduced
line luminosities? Unfortunately, the
only Local Group carbon sequence WR stars observed at metallicities below
the LMC are  the WO stars Sand~1 (Sk~188) in the SMC (Kingsburgh 
et al. 1995) and DR1 in IC~1613 (Kingsburgh \& Barlow 1995). Smith
et al. (1990a) established a uniform  C\,{\sc iv} $\lambda$5808
line luminosity for LMC WC stars, which we now re-evaluate based upon
a larger sample,  and compared with low metallicity WO stars, 
plus Sand~2 in the LMC (Kingsburgh et al. 1995).

\begin{table}[h]
\caption{Mean line luminosities  ($\pm$ standard deviations)
for the strongest
optical lines in single WC and WO stars in the Magellanic Clouds and 
IC~1613   (units of 10$^{35}$  erg/s), assuming distances of 49\,kpc, 
62\,kpc and 660\,kpc to the LMC, SMC and IC~1613, 
respectively.  For comparison, Smith et al. 
(1990a) adopted 3.2$\times 10^{36}$ erg/s for
C\,{\sc iv} $\lambda$5808 based on 5 LMC WC4 stars. 
Values in parenthesis additionally 
include WC+O binaries in the LMC and Sand~1 (WO+O) in the SMC.}
\label{average_wc}
\begin{tabular}{
l@{\hspace{3mm}}
l@{\hspace{2mm}}
l@{\hspace{3mm}}
l@{\hspace{2mm}}
l@{\hspace{3mm}}
l}
\hline
Line & WC4 & WO & WO \\
     & LMC & LMC & SMC/IC1613 \\
Number  & 7 (17) & 1 & 1 (2) \\
\hline
O\,{\sc iv} $\lambda$3411   &  11.6$\pm$4.6  & 9.1 & 13.3\\
                                       & --  & & (27.7$\pm$20.4)\\
O\,{\sc vi} $\lambda$3818   & 1.6$\pm$0.6  &  5.5 & 15.1\\
                                       & -- &  & (20.0$\pm$6.7) \\
C\,{\sc iii} $\lambda$4650/He\,{\sc ii} $\lambda$4686   
                                      &  49.2$\pm$22.7  &  5.2 & 8.6\\    
                                        & (48.8$\pm$20.9) & & (12.2$\pm$5.0)\\     
C\,{\sc iv} $\lambda$5808   &   33.6$\pm$16.1   & 10.2 &   8.6 
\\
                                       &(32.5$\pm$15.8) & & (11.3$\pm$5.0) \\
\hline
\end{tabular}
\end{table}

Spectrophotometry of 17 LMC WC stars were obtained with the Mt 
Stromlo 2.3m Dual Beam Spectrograph (DBS) in Dec 1997 
(see Crowther et al. 2002 for details). For the 9
stars in common with Torres-Dodgen \& Massey (1988), 
we find satisfactory agreement between
synthetic magnitudes and colours, with average differences of $\Delta v
=v_{\rm DBS}- v_{\rm TM88} = -0.05\pm0.05$ mag and $\Delta(b-v) =
(b-v)_{\rm DBS} - (b-v)_{\rm TM88} = +0.04\pm0.14$ mag, respectively. 
Seven of our sample are single according to Bartzakos et al. 
(2001), i.e. the  six studied by Crowther et al. (2002) plus MGWR5  
(Morgan \&  Good  1985, alias  BAT99--121).
Reddenings are either from Crowther et al. (2002), $(b-v)_{0}=-0.28$ mag 
for cases in which the WC star dominates the optical spectrum
i.e. BAT99-121 and BAT99-70 (Brey~62), or 
$(b-v)_{0}=-0.32$ mag otherwise. Smith et al. (1990a) assumed 
$(b-v)_{0}=-0.30$ mag for all LMC WC stars, regardless of binarity. 
For the 7 stars in common, we find $\Delta E_{B - V} = E_{B-V}({\rm
this~study}) - E_{B-V}({\rm S90}) = -0.04\pm0.08$~mag. 
Photometric properties are presented in Table A2 in the Appendix. 

Average WC (LMC) and WO (LMC, SMC, IC~1613) line luminosities for the
principal optical features are presented in Table~\ref{average_wc}. 
Our results support the conclusions of Smith et al. (1990a) that the
average C\,{\sc iv} $\lambda$5808 line luminosity of LMC WC stars is
3.2$\times 10^{36}$ erg/s, using an increased sample. Despite the small
number statistics, C\,{\sc iv} $\lambda$5808 line luminosities of WO
stars are uniformly a factor of $\sim$3 times lower than those of LMC WC
stars.  

Fig.~\ref{wc} compares the C\,{\sc iv} equivalent widths and FWHM
versus line luminosity, and includes an estimate of the line dilution for
the multiple system BAT99-10 for which spatially resolved HST/FOS
spectroscopy of the WC4 component has been obtained by Walborn et al.
(1999, their HD32228-2). The figure further illustrates that WO stars in 
all low metallicity environments possess lower line luminosities than LMC 
WC stars. Ultimately, the robustness of these results depends on the
universality of the empirical trend from WC to WO at low metallicity (see
Sect.~\ref{models_wc}).

%
%

\begin{figure}[h]
 \centerline{\psfig{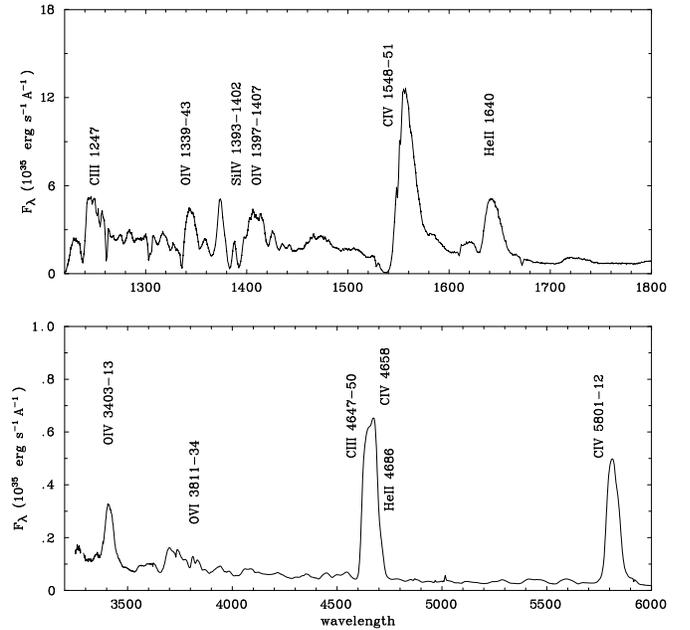}} 
\caption{`Generic' UV and optical spectrum of an early WC star in 
the LMC.}\label{generic-wc} 
\end{figure}

As with the WN sequence, we have constructed a `generic' LMC-metallicity
early WC spectrum from our 7 single WC4 stars, which is presented in  
Fig.~\ref{generic-wc}. This has potential use in synthesising the 
WR  bumps in  moderately metal-poor galaxies (Hadfield \&  Crowther, in 
preparation).

\subsection{C\,{\sc iv} $\lambda$1550 in WC stars}\label{1548}

We have derived the average C\,{\sc iv} $\lambda$1550 emission
line flux of six LMC WC4 stars from HST/FOS observations presented in
Gr\"{a}fener et al. (1998) and Crowther et al. (2002), revealing F(C\,{\sc
iv} $\lambda$1550)/F(C\,{\sc iv} $\lambda$5808)
= 6.0$\pm$0.6, and F(He\,{\sc ii} $\lambda$1640)/F(C\,{\sc iv}
$\lambda$1550) = 0.43 $\pm$0.11. These values agree with
Schaerer \& Vacca (1998) who obtained F(He\,{\sc ii}
$\lambda$1640)/F(C\,{\sc iv} $\lambda$5808) = 2.14$\pm$1.09 from
16 (mostly binary) WC4 stars. Sand~2 (WO) reveals similar values, with
F(C\,{\sc iv} $\lambda$1550)/F(C\,{\sc iv}
$\lambda$5808) $\sim$6.2 and F(He\,{\sc ii}
$\lambda$1640)/F(C\,{\sc iv} $\lambda$1550) $\sim$0.25. {\bf
Fig.~\ref{generic-wc} includes a generic ultraviolet LMC early-type WC
spectrum, which } was obtained from the HST/FOS observations, scaled
relative to the generic optical spectrum, such that F(He\,{\sc ii}
$\lambda$1640)/F(C\,{\sc iv} $\lambda$5808) = 2.6.

\section{Role of metallicity-dependent WR winds}\label{models}

We have demonstrated that WR stars at sub-LMC metallicities possess
reduced line luminosities, by factors of $\leq$3 (WC sequence) to 
4--5 (WN sequence), with respect to higher metallicity 
counterparts of the same spectral subclass. 


Up until recently, the winds of WR stars were assumed to be metallicity
independent (Langer 1989). Crowther et al. (2002) claimed a metallicity
dependence of WC winds from an analysis of LMC and Milky Way stars.
Recently, Gr\"{a}fener \& Hamann (2005) have argued that line driving of
WR winds is dominated by Fe-peak elements, whilst Vink \& de Koter (2005)
argue for $\dot{M} \propto Z^{0.86}$ for cool WN stars with $10^{-3} \leq
Z/Z_{\odot} \leq 1$, where $Z$ is the initial
heavy metal content. For cool WC stars Vink \& de Koter (2005) propose
 $\dot{M} \propto Z^{0.66}$ for $0.1 \leq Z/Z_{\odot} \leq$1, and 
$\dot{M} \propto Z^{0.35}$ for $10^{-3} \leq Z/Z_{\odot} \leq$0.1.

\begin{table}[h]
\caption[]{
Physical parameters for the Solar metallicity late- and early-type
WN  models (WNL-1 adapted from Herald et al. 2001, WNE-1 adapted from Morris et al. 2004), 
together with the low metallicity models (WNL-2 and WNE-2), in which the  metal content has 
been uniformly reduced to 1/50 $Z_{\odot}$, and the  mass-loss  rate reduced by a factor of 50$^{0.7}$. 
Clumping is incorporated via a volume filling factor $f$=0.1 (Hillier 1991). Nitrogen and iron mass fractions 
are indicated. Line luminosities of He\,{\sc ii} $\lambda$1640 and $\lambda$4686 are listed (the 1640 line
in the WNL-1 model is a blend). $Q_{0}, Q_{1}, Q_{2}$ are
the number of ionizing photons shortward of the H$^{0}$, He$^{0}$ and He$^{+}$ 
edges.}
\label{wn_models}
\begin{tabular}{lllll}
\hline
Model & WNL-1 & WNL-2 & WNE-1 & WNE-2 \\
\hline
$T_{\ast}$(kK) & 45 & 45 & 90 & 90 \\
$\log L/L_{\odot}$ & 5.60 & 5.60 & 5.60 & 5.60 \\
$\log \dot{M}/M_{\odot}$yr$^{-1}$ & --4.5 & --5.7 & --4.5 & --5.7 \\
$v_{\infty}$(km/s) & 850 & 850 & 1900 & 1900 \\
X(H) & 15\% & 15\% & 0 & 0 \\
X(N) & 1.2\% & 0.02\% & 1.2\% & 0.02\% \\
X(Fe) & 0.12\% & 0.0024\% & 0.12\% & 0.0024\% \\
$M_{v}$(mag) & --6.1  & --5.3    & --4.8 & --3.2 \\ 
$L_{\rm 1640}$ (erg/s) &-&2.0$\times 10^{36}$& 2.0$\times 10^{37}$ & 1.1$\times 10^{36}$ \\
$L_{\rm 4686}$ (erg/s) &1.6$\times 10^{36}$&2.2$\times 10^{35}$& 2.0$\times 10^{36}$ & 1.1$\times 10^{35}$ \\
$\log Q_{0}$ &49.4&49.4& 49.5 & 49.5 \\
$\log Q_{1}$ &48.4&48.9& 49.15& 49.25\\
$\log Q_{2}$ &..&..& 40.6 & 48.1 \\
Sp Type & WN8 & WN5ha & WN4 & WN2 \\
\hline
\end{tabular}
\end{table}

 In the following, we shall assume that WR winds scale with metallicity 
as $\dot{M}_{\rm WN}$ $\propto Z^{0.7}$ and $\dot{M}_{\rm WC}$ 
 $\propto  Z^{0.5}$,
in part based on recent empirical results (Crowther 2006), for
$10^{-2} \leq Z/Z_{\odot} \leq 1$. We  now investigate
how line luminosities and ionizing spectra from WN and WC stars vary with 
metallicity under such assumptions.

\subsection{WN stars}\label{models_wn}

We have selected HD~96548 (WN8) and HD~50896 (WN4) as
representative  single strong-lined Galactic late and early-type WN 
stars, for which 
spectroscopic analyses were performed by Herald et al. (2001) and Morris et al. (2004)
based on the non-LTE, line blanketed model atmosphere code CMFGEN (Hillier \& Miller
1998), which solves the radiative transfer equation in the co-moving frame, subject to 
the additional constraints of radiative and statistical equilibrium. 
New  calculations are performed for the same ions considered by  Herald et al. 
(2001) and Morris et al. (2004), with some subtle differences in atomic 
data.

\begin{figure}[h]
 \centerline{\psfig{figure=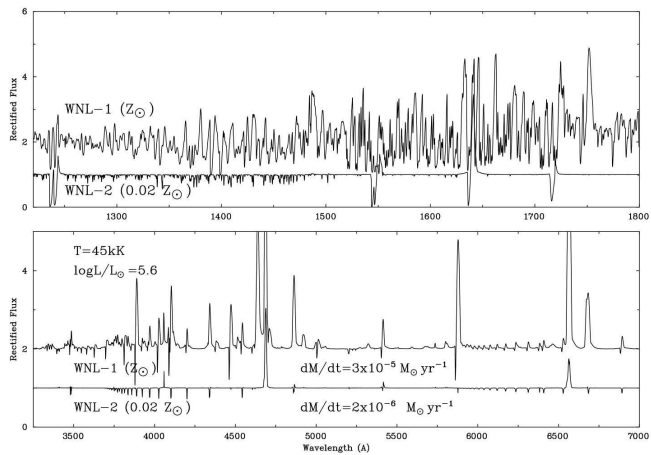,width=9.5cm,angle=-90.}} 
\caption{Comparison between late-type  WN CMFGEN models in which parameters are 
fixed,  except heavy metal abundances differ by a factor of 50, and mass-loss 
rates differ by a factor of $50^{0.7}$. The high mass-loss model closely matches the observed
spectrum of HD~96548 (WN8, Herald et al. 2001) whilst the  low mass-loss model may be representative of 
a  low temperature WN star in I\,Z~18, with a  WN5ha spectral type.} 
\label{wnl_model}
\end{figure}

\begin{figure}[h]
 \centerline{\psfig{figure=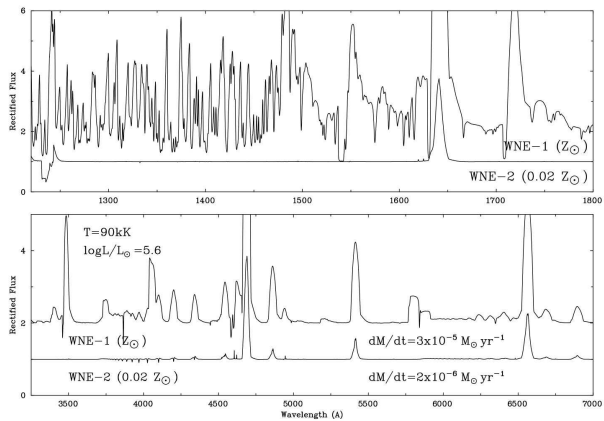,width=8.8cm,angle=-90.}} 
\caption{Comparison between early-type WN CMFGEN models in which parameters are 
fixed,  except heavy metal abundances differ by a factor of 50, and mass-loss rates differ by a 
factor of $50^{0.7}$.  The high mass-loss model closely matches the observed
spectrum of HD~50896 (WN4, Morris et al. 2004), whilst  the  low mass-loss model
may be representative of a high temperature WN star in I\,Z~18, with WN2 spectral type.} 
\label{wne_model}
\end{figure}

We present the rectified synthetic UV and optical  Solar-metallicity 
models of late- (hereafter WNL-1) and  early-type (hereafter 
WNE-1) 
WN stars in Figs.~\ref{wnl_model}--\ref{wne_model}, together with identical 
atmospheric models, except that the heavy metal  content has 
been reduced from $Z_{\odot}$ to 1/50 $Z_{\odot}$ and the mass-loss rate has 
been reduced by a factor  of 50$^{0.7}$ (hereafter WNL-2 and WNE-2).  Although
metal-poor WNE stars may be rich in hydrogen (e.g. Foellmi et al. 2003a) 
the actual
hydrogen content does not significantly affect the ionizing flux or UV/optical spectrum,
beyond the strength of the Balmer lines.

The low metallicity late-type WN model reveals a  weak emission
line spectrum, with primarily He\,{\sc ii} $\lambda$4686 and N\,{\sc iv} $\lambda$4058 
seen in the blue visual, such that the Solar-metallicity WN8 model has shifted
to a weak-lined WN5ha subtype at low metallicity  based on the Smith
et al. (1996) classification scheme. Similarly, the low 
metallicity early-type WN model effectively displays a pure He\,{\sc ii} 
emission line spectrum in the visual,
such that the Solar-metallicity strong-lined WN4 model has shifted to a 
weak-lined WN2 subtype. In the
ultraviolet, weak C and N features are seen in both cases. The principal physical
parameters of each model are presented in Table~\ref{wn_models}.

\begin{figure}[h]
 \centerline{\psfig{figure=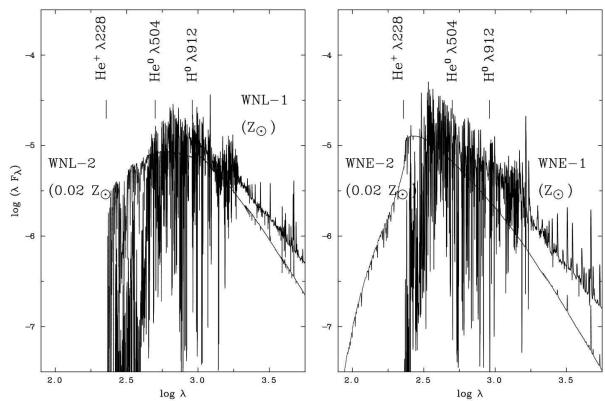,width=8.8cm,angle=-90.}}
\caption{Comparison between predicted emergent spectral energy
distributions (erg\,cm$^{-2}$\,s$^{-1}$ at 1\,kpc) of late-type (left
panel) and early-WN (right panel) CMFGEN models, with identical parameters
except that heavy metal abundances differ by a factor of 50, and mass-loss
rates differ by a factor of 50$^{0.7} \sim 15$, such that the reduced
mass-loss models (dotted) display harder flux distributions than the high
mass-loss cases (solid).} 
\label{wn_sed} 
\end{figure}

The He\,{\sc ii} $\lambda$4686 line luminosity of model WNE-2 is severely
reduced relative to WNE-1, due to the combination of reduced line
equivalent width (factor of $\sim$4) and reduced optical continuum flux
(factor of $\sim$5). The high wind density of the WNE-1 model prevents emission
shortward of the He$^{+}$ 228\AA\ edge, which is re-radiated at near-UV
and optical wavelengths, as previously discussed by Schmutz et al. (1992)
and Smith et al. (2002). In contrast, the low wind density of the low
metallicity model WNE-2 produces a hard extreme UV flux distribution, as
indicated in Fig.~\ref{wn_sed} (see also Table~\ref{wn_models}).
Note that the empirical WN flux ratio F(He\,{\sc ii} 
$\lambda$1640)/F(He\,{\sc ii} $\lambda$4686) 
= 10 (Sect.~\ref{1640}) is also supported at low metallicity.

\subsection{WC stars}\label{models_wc}

We have carried out similar calculation for WC stars, based upon models
for the single WC9 star HD~164270 in the Milky Way from Crowther et
al. (2006, hereafter WCL-1) and the WC4 star HD~37026 (BAT99-52) in the
LMC from Crowther et al. (2002, hereafter WCE-1). Once again, we have
obtained CMFGEN model atmospheres for the same ions considered by Crowther
et al. (2002, 2006) except for some subtle differences in atomic data.
These calculations were repeated except that the heavy elemental
abundances (beyond Ne) have been reduced from 1$Z_{\odot}$ (or 1/2
$Z_{\odot}$) to 1/50 $Z_{\odot}$ and reduced the mass-loss rate by a
factor of 50$^{0.5}\sim 7$ or 25$^{0.5}\sim 5$, in order to mimic late-
and early-type WC stars in I\,Zw~18 (hereafter WCL-2 and WCE-2).

\begin{table}[h]
\caption[]{
Physical parameters for the Solar metallicity late-type and LMC metallicity
early-type WC  models (WCL-1 adapted from Crowther et al. 2006, WCE-1 
adapted from Crowther et al. 2002, 
together with the low metallicity models (WCL-2 and WCE-2), in which the  metal content has 
been uniformly reduced to 1/50 $Z_{\odot}$, and the  mass-loss  rate reduced by factors of 50$^{0.5}$
(Solar to I\,Zw~18) and 25$^{0.5}$ (LMC to I\,Zw~18). Clumping is incorporated in all models 
via a volume filling factor $f$=0.1. Carbon, oxygen and iron mass fractions are indicated.
Line luminosities of C\,{\sc iv} $\lambda$1550 (blended in 
WCL-1), $\lambda$5808 (blended with C\,{\sc iii} $\lambda$5826 in WCL-1)
and C\,{\sc iii} $\lambda$5696, $\lambda$4650/He\,{\sc ii} 
$\lambda$4686, 
are listed. $Q_{0}, Q_{1}, Q_{2}$ are the number of 
ionizing photons shortward of the H$^{0}$, He$^{0}$ and He$^{+}$ edges.}
\label{wc_models}
\begin{tabular}{lllll}
\hline
Model & WCL-1 & WCL-2 & WCE-1 & WCE-2 \\
\hline
$T_{\ast}$(kK) & 50 & 50 & 100 & 100 \\
$\log L/L_{\odot}$ & 4.90 & 4.90 & 5.65 & 5.65 \\
$\log \dot{M}/M_{\odot}$yr$^{-1}$ & --5.0 & --5.8 &  --4.5 & --5.2 \\
$v_{\infty}$(km/s) & 1050 & 1050 & 2900 & 2900 \\
X(C) & 35\% & 35\% & 44\% & 44\% \\
X(O) & 5\% & 5\% & 9\%      & 9\%        \\
X(Fe) & 0.12\% & 0.024\% &  0.06\% & 0.0024\%\\
$M_{v}$(mag) & --4.6 & --3.4 & --4.8 & --3.4 \\ 
$L_{\rm 1550}$ (erg/s) &..&7.2:$\times 10^{35}$& 2.5$\times 10^{37}$ & 5.1$\times 10^{36}$ \\
$L_{\rm 4650}$ (erg/s) &3.6$\times 10^{35}$& 2.6$\times 10^{35}$& 7.0$\times 10^{36}$ & 3.5$\times 10^{35}$ \\
$L_{\rm 5696}$ (erg/s)     &5.6$\times 10^{35}$& 2.2$\times 10^{34}$  & --  & --\\
$L_{\rm 5808}$ (erg/s) &1.1$\times 10^{35}$ & 3.9$\times 10^{34}$& 2.5$\times 10^{36}$ & 4.0$\times 10^{35}$ \\
$\log Q_{0}$ & 48.6 &48.8 & 49.5 & 49.5 \\
$\log Q_{1}$ & 40.4 &48.0 & 49.0 & 49.2\\
$\log Q_{2}$ & .. &..& ..& 47.5 \\
Sp Type & WC9 & WC7 & WC4 & WC4 \\
\hline
\end{tabular}
\end{table}

The reduced metallicity synthetic spectra reveal weak-lined WC spectral
types, since the assumed C and O abundances remain high in the reduced
metallicity models: WC7 in the case of WCL-2, and WC4 in the case of
WCE-2 following the classification schemes of Crowther et al. (1998)
or Smith et al. (1990b).  
O\,{\sc iv} $\lambda$3411 and O\,{\sc v}
$\lambda$5592 are strong in WCE-2, with O\,{\sc vi}
$\lambda$3818 weak.  Note a modest increase in temperature to
110kK is sufficient to produce strong O\,{\sc vi} $\lambda$3818
emission via a switch in the oxygen ionization balance, i.e. a WO subtype.

\begin{figure}[h]
 \centerline{\psfig{figure=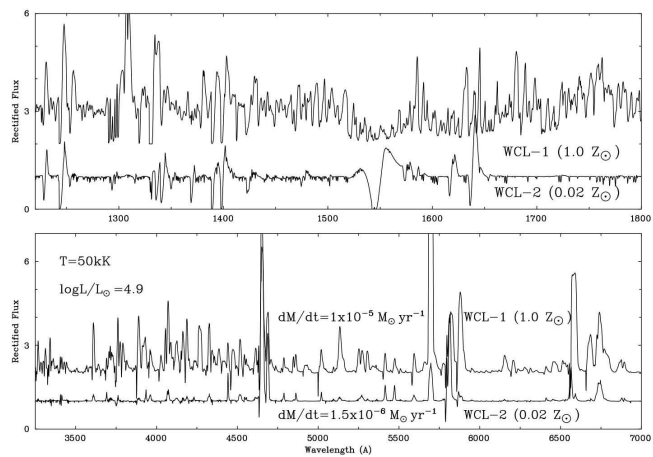,width=9.5cm,angle=-90.}} 
\caption{Comparison between late-type WC CMFGEN models in which parameters are fixed, 
except the heavy metal abundances differ by a factor of 50, and mass-loss rates differ by 
a  factor of 50$^{0.5} \sim 7$.  The high mass-loss model closely matches the observed
spectrum of HD~164270 (WC9, Crowther et al. 2005)  whilst the  low mass-loss model 
may be representative of a late-type WC star in I\,Z~18, with spectral type WC7.} 
\label{wcl_model}
\end{figure}

The low wind density WC models reveal spectral morphologies quite distinct
from the high wind density cases, despite identical (C+O) abundances.
Efficient metal wind cooling in the WCL-1 and WCE-1 models (Hillier 1989) produces 
strong C\,{\sc iii} lines. For example, C\,{\sc iii} $\lambda$4650 
dominates the $\lambda$4650 feature in the WCE-1 model, with F(C\,{\sc iii} 
$\lambda$4650) $\geq$ F(He\,{\sc ii} $\lambda$4686) $\geq$ 
F(C\,{\sc iv} 
$\lambda$4659). In contrast, cooling is greatly reduced in the low metallicity 
models, such that negligible C\,{\sc iii} emission is now predicted
in the WCE-2 
model, with  F(He\,{\sc ii} $\lambda$4686) $\sim$ F(C\,{\sc iv} $\lambda$4659)
contributors to the $\lambda$4670 feature.  
Consequently, typical line ratios of individual
WC stars at LMC or Solar metallicity -- e.g. C\,{\sc iii}
$\lambda$4650/C\,{\sc iv} $\lambda$5805 $\sim$1.6 for WC4 stars 
(Smith et al. 1990a) -- do not necessarily hold for WC stars in 
metal-poor environments.
Indeed, one anticipates the need for a new system of WR spectral classification 
at low metallicity, based upon a set of standard stars, as in the Solar metallicity case.

\begin{figure}[h]
 \centerline{\psfig{figure=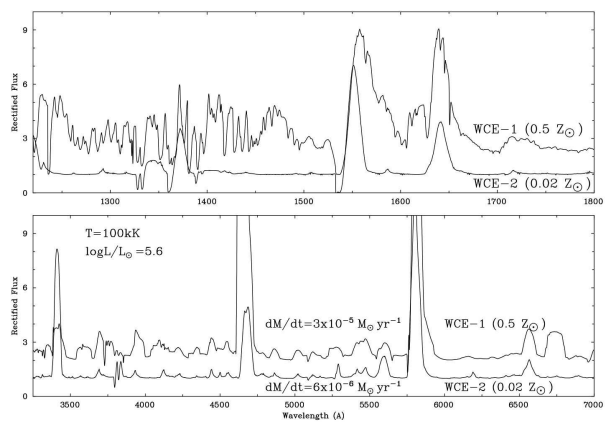,width=8.8cm,angle=-90.}} 
\caption{Comparison between early-WC CMFGEN models in which parameters are fixed, 
except heavy metal abundances differ by a factor of 25, and mass-loss rates differ by 
a  factor of 25$^{0.5} \sim 5$.  The high mass-loss model closely matches the observed
spectrum of the LMC WC4 star HD~37026 (Crowther et al. 2002), whilst 
the  low mass-loss model may be representative of an early-type WC
star in I\,Z~18, with WC4 or WO spectral type.} 
\label{wce_model}
\end{figure}

\begin{figure}[h]
 \centerline{\psfig{figure=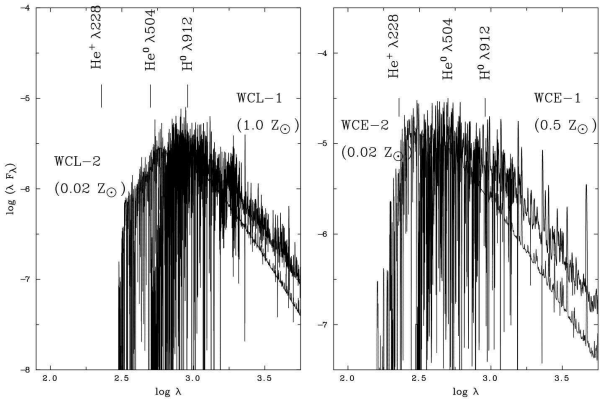,width=8.8cm,angle=-90.}}
\caption{Comparison between predicted emergent spectral energy
distributions (erg\,cm$^{-2}$\,s$^{-1}$ at 1\,kpc) of late-type (left
panel) and early-type (right panel) WC CMFGEN models, with identical
parameters except that heavy metal abundances differ by a factor of
25--50, and mass-loss rates differ by factors of Z$^{0.5}$, such that the
reduced mass-loss models (dotted) display harder flux distributions than
the high mass-loss cases (solid).} \label{wc_sed} \end{figure}

Similar arguments to those discussed above for WN stars cause the
C\,{\sc iv} $\lambda$5808 line luminosity of the low metallicity
models to be reduced by factors of 3--6 relative to the Solar/LMC models,
due to the combination of reduced line equivalent widths and reduced
optical continuum fluxes.  As before, 
the low metallicity, low wind density models  predict much 
harder ionizing
flux distributions than their high metallicity, high wind density
counterparts from an observer's perspective as indicated in 
Fig.~\ref{wc_sed} (see also Table~\ref{wc_models}).

\section{Discussion}\label{discussion}

We have demonstrated that line luminosities of early--mid WN stars in the
LMC exceed those in the SMC by $\sim$4--5. If this is supported for other
metal-poor galaxies, one would need to apply such a corrective factor when
determining WR populations at low metallicity. Unfortunately, 
within the Local Group only NGC~6822 and IC~10 --  with log (O/H)+12 
$\sim$ 8.25 (Pagel et  al. 1980; Garnett 1990) -- possess metallicities 
below that of the LMC. To date, only 4 and 11 WN stars have been confirmed 
in these respective galaxies (Massey \& Johnson 1998; Crowther et al. 
2003) whilst spectrophotometry is not yet available, preventing robust 
line flux comparisons since reliable reddening corrections rely upon 
accurate colours of isolated WR stars.

We propose that a correction factor may need to be applied when estimating
WR populations from observations of unresolved clusters/galaxies at
sub-LMC metallicities, where {\it adopted} values from metal-rich WR
calibrations may greatly exceed those of individual stars within those
galaxies. In the extreme case of I\,Zw~18, our test calculations 
suggest factors of $\sim$5--20 may be appropriate.

\subsection{WR population of I\,Zw~18 from optical studies}

Izotov et al. (1997) and Legrand et al. (1997) 
presented observations of I\,Zw~18-NW in which broad blue
(C\,{\sc iii} $\lambda$4650/C\,{\sc iv} $\lambda$4658/He\,{\sc ii} 
$\lambda$4686, FWHM$\sim$70\AA)
 and yellow (C\,{\sc iv} $\lambda$5808, FWHM$\sim$50\AA)  
emission features were observed, together with nebular He\,{\sc ii}
$\lambda$4686 emission.  Izotov et al. derived blue and yellow line
luminosities of 4.8$\times 10^{37}$ erg/s and 2.1$\times 10^{37}$ erg/s,
respectively, for a revised distance of $\sim$14.1\,Mpc (Izotov \& Thuan
2004). Applying  our own LMC WC calibration or that from Smith et al. 
(1990a) to the yellow feature would require $\sim$7 equivalent WC4 stars 
~(De  Mello et al. 1998). In contrast, with a typical low metallicity WC 
star  contributing a factor of $\sim$5 less C\,{\sc iv} 
$\lambda$5808  flux, we suggest a  much larger WC population of 
$\geq$30 
may be necessary to explain the 
observed line flux in I\,Zw~18.  Depending upon individual temperatures, 
these stars would display either a weak-lined early WC, or a WO spectrum.

\begin{figure}[h]
 \centerline{\psfig{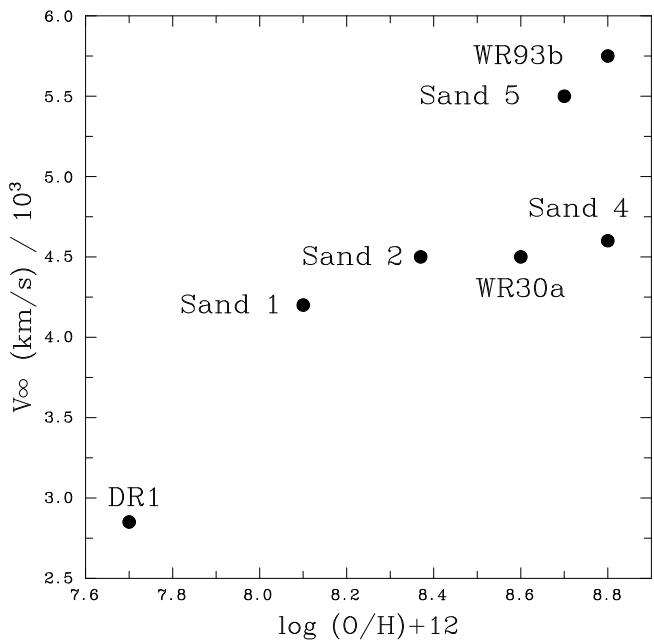}} 
\caption{Comparison between wind velocities of all known Local Group WO
stars, from Kingsburgh et al. (1995), Kingsburgh \& Barlow (1995) and 
Drew et al. (2004). For Milky Way WO stars we have adopted $\log$(O/H)+12=8.6 for
WR30a which lies beyond the Solar circle, 8.7 for Sand~5 which lies at the Solar
circle and 8.8 for Sand~4 and WR93b which lie within the Solar circle.}
\label{wo}
\end{figure}

Legrand et al. (1997) noted that the observed line width of the C\,{\sc
iv} feature more closely matches that of LMC WC4 stars than WO stars 
(recall Fig.~\ref{wc}). However, WO stars are known to display decreasing 
wind velocities at lower metallicity, as demonstrated in Fig.~\ref{wo}.  
Consequently, one might expect low metallicity WO stars to have unusually 
narrow lines with respect to their metal-rich counterparts.

Izotov et al. noted that the C\,{\sc iii-iv} $\lambda$4650 flux 
far exceeded that expected from the number of WC stars inferred from 
C\,{\sc iv}
$\lambda$5808, assuming they were typical of LMC WC4 stars, which they
exclusively attributed to  He\,{\sc ii} $\lambda$4686 in 
late-type WN stars.  If WC stars in I\,Zw~18 mimic those of the LMC
one would expect C\,{\sc iii-iv} $\lambda$4650/C\,{\sc iv} $\lambda$5808
=1.5--1.6 (Table~\ref{average_wc}; Smith et al. 1990a) so these would
provide $\sim$2/3 of the C\,{\sc iii-iv} $\lambda$4650 flux observed
by Izotov et al. (1997). The remainder could be attributed to $\sim$10 
WN5--6 stars, or  $\sim$20 WN7--9 (or WN2--4) LMC-like stars 
(Table~\ref{average_wn}; see  also De Mello et al. 1998).

We have shown that low metallicity WC stars are likely to possess rather
different ratios of C\,{\sc iii} $\lambda$4650 to C\,{\sc iv}
$\lambda$5808 line fluxes from near-Solar counterparts, so we {\bf advise}
caution when indirectly inferring WR populations at extremely low
metallicities in this way.  Observationally, it is challenging to 
establish the presence of WN stars in the Izotov et al.  (1997) dataset
since the $\lambda$4650 feature is much broader than in late-type WN
stars, with FWHM $\sim$70\AA. Late-type LMC metallicity WN stars possess
FWHM$\sim$20\AA, at comparable spectral resolution (Fig.~\ref{wn}).

The difficulty in  spectroscopically identifying WN stars in I\,Zw~18 is
almost certainly due to the extremely low He\,{\sc ii} $\lambda$4686 line
luminosity of individual WN stars, owing to the steeper metallicity
dependence of their winds relative to WC stars (Vink \& de Koter 2005).  
Since late-type WN stars positively shy away from low metallicity
environments (recall Fig.~\ref{wnl_model}), line luminosities of
individual mid-type or early-type WN stars in I\,Zw~18 may be a factor of
$\sim$10 times smaller than for  LMC late-type WN stars according
to Schaerer \& Vacca (1998). Indeed, if we assume $\sim$1/3 of the 
C\,{\sc iii-iv} $\lambda$4650 flux observed by Izotov et al. (1997)
is due  to SMC early-type 
WN stars, we would require 100--300 WN stars, depending upon whether
we adopt average values from Table~\ref{average_wn} including or 
excluding the WN+O binaries. Consequently, large numbers of WN stars 
would be required for their  spectroscopic detection  via broad 
He\,{\sc ii} $\lambda$1640 or $\lambda$4686 emission. Alternatively,   
their presence  in smaller numbers may be seen  indirectly via strong 
nebular  He\,{\sc ii} $\lambda$4686 emission,  which {\it is} indeed 
observed in I\,Zw~18.

\subsection{WR population of I\,Zw~18 from UV studies}

Brown et al. (2002) scanned the NW part of I\,Zw~18 with HST/STIS, 
revealing two  clusters, exhibiting strong C\,{\sc iv} 
$\lambda$1550 and He\,{\sc ii} $\lambda$1640 emission.
 Adjusting their fluxes to a distance of 14.1\,Mpc (Izotov \& Thuan 
2004) indicates He\,{\sc ii} $\lambda$1640 luminosities of 
L$_{\rm 1640}$ = 3.0 $\times 10^{37}$ erg/s and 4.0 
$\times 10^{37}$ erg/s, respectively. Brown et al. applied the
Schaerer \& Vacca (1998) He\,{\sc ii} $\lambda$1640 calibration of a 
representative {\it Milky Way} WC5 star,  implying 6 and 8 WC stars, 
respectively, with   N(WC)/N(O)$\sim$0.2 in the former cluster. 

A  similar exercise for the (distance adjusted)  C\,{\sc iv}
$\lambda$1550 luminosities of L$_{\rm 
1550}$ = 6.9 $\times 10^{37}$ erg/s and 2.8 $\times 10^{37}$ erg/s 
would lead to 3--4 and 1--2 stars, 
respectively, according to Sect.~\ref{1548} assuming 
representative LMC-type WC4 stars. The reduced numbers
with respect to Brown et al. (2002) is due to the higher 
line luminosities of LMC WC4 stars relative to Milky Way WC5 stars, plus 
the low $\lambda$1550/$\lambda$1640 ratio of $\sim$0.7 for the second 
cluster, suggesting a primary contribution by WN stars rather than WC 
stars.

If we instead assume a representative C\,{\sc iv} $\lambda$1550 line
luminosity of 4$\times 10^{36}$ erg/s for a single WC star in
I\,Zw~18, i.e. 5 times lower than typical LMC WC4 stars, we would require
$\sim$17 and $\sim$7 weak-lined WC stars for the two clusters. Brown
et al. remarked upon their unusually high N(WR)/N(O) populations, which
would be exacerbated if larger WC populations are inferred at reduced
metallicities.

\subsection{Challenges for evolutionary models of single massive 
stars at low metallicity}

The potential presence of much greater numbers of WR stars in I\,Zw~18
than is currently appreciated naturally causes problems for evolutionary
models of  single massive stars at low metallicity.  Non-rotating, 
high mass-loss evolutionary models were calculated by de Mello et al. 
(1998), revealing progression through to the WN and WC phases for stars of 
initial mass $\sim 90M_{\odot}$ and 120$M_{\odot}$, respectively. For an 
instantaneous burst with a Salpeter IMF and an upper mass limit of 
$\sim150M_{\odot}$ the maximum N(WR)/N(O) ratio is $\sim$0.02, with WN 
stars dominating the WR population, i.e. N(WC)/N(O)$\leq$0.003. In the 
case of a WR population 
with $\sim$30 WC or WO stars, and potentially $\sim$200 WN stars, 
one would obtain  N(WC)/N(O)$\sim$0.02   and N(WN)/N(O)$\sim$0.1  in 
I\,Zw~18, based  upon the $\sim$2000 O star content from Izotov et al. 
(1997, again  adjusted to a distance of  14.1\,Mpc) greatly exceeding 
evolutionary  predictions for single stars. 
Further comparison awaits the calculation of evolutionary models for 
single stars
at very low metallicity including rotation and contemporary mass-loss
rates.  

\subsection{Circumstellar environment of long duration GRBs}

Within the past decade, several long duration ($\geq$ 2s) GRBs have been 
positively  identified with Type Ic core-collapse SN
(Galama et al. 1998; Stanek et al. 2003), supporting the 
collapsar model of MacFadyen \& Woosley (1999) involving Wolf-Rayet
stars. The ejecta strongly interacts with the circumstellar material, probing 
the immediate vicinity of the GRB itself, thus  providing information 
on the progenitor (Li \& Chevalier 2003, van Marle et al. 2005). 

If WR stars possess metallicity-dependent winds, one would potentially expect 
rather different environments for the afterglows of long-duration 
GRBs, that were dependent  upon the  metallicity of the host galaxy. WN 
stars in a  galaxy of 1/100$Z_{\odot}$ may possess wind densities a factor of 
$\sim$25 times lower than those in the Milky Way. In general, the metallicity 
dependence of wind velocities for WR stars is unclear, although amongst
carbon sequence WR star, lower velocity winds are seen in 
WO stars from metal-poor environments (Fig.~\ref{wo}). 

Overall, the immediate environment of GRBs that involve Wolf-Rayet
precursors may differ substantially from those of Solar metallicity WR
stars (Eldridge et al. 2005). Indeed, the host galaxies of high-redshift
GRBs tend to be rather metal-poor, from medium to high resolution
spectroscopy obtained immediately  after the burst.  For example,
Vreewwijk et al. (2004)  suggest 1/20$Z_{\odot}$ for the host galaxy of
GRB 030323 at $z$=3.37 and Chen et al. (2005) conclude 1/100$Z_{\odot}$
for the host galaxy of GRB 050730 at $z$=3.97.

\section{Summary}\label{summary}

We have demonstrated empirically that individual WN and WC stars at SMC
metallicities possess lower optical line luminosities than those in the
LMC (or Milky Way), which currently represent the standard calibrations
for WR populations in external galaxies (Schaerer \& Vacca 1998). Reduced
optical line luminosities at lower metallicities {\it naturally} follow if
the strength of WR winds depends upon metallicity, as recently proposed
(Crowther et al. 2002; Vink \& de Koter 2005), due to the combination of
smaller line equivalent widths and lower optical continuum levels.

Wolf-Rayet stars with weak winds are capable of producing significant
He\,{\sc ii} Lyman continuum photons (Schmutz et al. 1992; Smith et al.
2002), which we attribute to the origin of nebular He\,{\sc ii}
$\lambda$4686 in low metallicity galaxies. Application to I\,Zw~18
suggests that WC stars are present in greater numbers than has been
previously suggested, and that WN stars are extremely difficult to detect,
since their winds appear to depend more sensitively upon metallicity than
WC stars.  An increased number of WR stars at low metallicity causes
severe problems with evolutionary predictions for single stars.

Finally, reduced wind strengths from WR stars at low metallicities impacts
upon the immediate circumstellar environment of long-duration GRB
afterglows, particularly since the host galaxies of high-redshift GRBs
tend to be metal-poor.

\begin{acknowledgements} The majority of our NTT/EMMI observations were
carried out in service mode, courtesy of John Pritchard, to whom we are
grateful.  We wish to thank Yuri Izotov for providing us with his MMT 2D
spectrum of I\,Zw~18  and Tony Moffat for a comprehensive referee's report
which has helped to improve the original manuscript.
 Financial support was provided by the Royal Society
(PAC) and PPARC (LJH). \end{acknowledgements}

\begin{appendix}

\begin{table*}
\begin{flushleft}
{\bf{Table A1.}} 
Photometric properties of single WNE stars in the LMC and SMC
 based upon our NTT/EMMI spectrophotometry, including He\,{\sc ii} 
$\lambda$4686 line measurements.
Interstellar extinction  corrections follow a standard Seaton (1979) 
reddening  law.  Distances of 49\,kpc and 62\,kpc are adopted for the 
LMC and SMC, respectively (see text). We compare synthetic magnitudes with 
 Torres-Dodgen \& Massey (1988, TM88) and reddenings with Schmutz \& Vacca 
(1991, SV91) and  Massey \& Duffy (2001, MD01). Two entries are listed for 
SMC WR11, providing photometry for the composite WN+? system
and the observed He\,{\sc ii} $\lambda$4686 flux, plus approximate
photometry for the WN star, for which the resulting reddening has been used 
to  obtain the He\,{\sc ii} $\lambda$4686 line luminosity.\\
\end{flushleft}
\begin{tabular}{c@{\hspace{2mm}}c@{\hspace{2mm}}c@{\hspace{2mm}}l@{\hspace{1.5mm}}rrrrrccccc
@{\hspace{-1mm}}c@{\hspace{-1mm}}c}
\hline
\multicolumn{2}{c}{Star}&&Sp&\multicolumn{1}{c}{\it v}&\multicolumn{1}{c}{\it b-v}&\multicolumn{1}{c}{\it
v-r}&\multicolumn{1}{c}{\it$\Delta$v}&\multicolumn{1}{c}{\it$\Delta$(b-v)}&
FWHM&\multicolumn{1}{c}{W$_{\lambda}$}&$F_{4686}$&$E_{B-V}$&$E_{B-V}$ 
&$L_{4686}$\\
BAT99&Br&&Type&mag&mag&mag&\multicolumn{2}{c}{-- TM88 --}& \AA& \AA& erg 
s$^{-1}$cm$^{-2}$ &SV91&This work&erg\,s$^{-1}$\\
\hline
1&1&&WN3b&15.75&0.01&--0.03&--0.17&0.00&33&255&6.5$\times 10^{-13}$&0.16&0.29&5.2$\times 10^{35}$\\
3&3&&WN4b&14.63&0.01&--0.08&--0.20&0.03&33&236&1.7$\times 10^{-12}$&0.11&0.29&1.3$\times 10^{36}$\\
5&4&&WN2b&16.67&0.06&0.02&--0.21&--0.01&37&180&1.9$\times 10^{-13}$&0.37&0.37&2.0$\times 10^{35}$\\
7&6&&WN4b&13.81&--0.10&--0.09&--0.06&--0.04&60&481&7.7$\times 10^{-12}$&0.26&0.13&3.5$\times 10^{36}$\\
15&12&&WN4b&14.39&--0.03&--0.05&--0.22&0.04&35&312&3.0$\times 10^{-12}$&0.25&0.22&1.8$\times 10^{36}$\\
17&14&&WN4o&14.16&--0.11&--0.15&--0.21&0.03&30&140&1.7$\times 10^{-12}$&0.06&0.20&9.7$\times 10^{35}$\\
18&15&&WN3(h)&14.60&--0.17&--0.19&--0.12&--0.01&30&123&1.0$\times 10^{-12}$&0.09&0.14&4.7$\times 10^{35}$\\
23&-&&WN3(h)&16.98&0.31&0.16&-&-&30&93&6.4$\times 10^{-14}$&-&0.78&2.9$\times 10^{35}$\\
24&19&&WN4b&14.52&--0.04&--0.06&--0.07&--0.05&41&375&3.1$\times 10^{-12}$&0.02&0.19&1.7$\times 10^{36}$\\
25&19a&&WN4ha&15.10&--0.23&--0.23&-&-&28&40&2.0$\times 10^{-13}$&-&0.08&7.4$\times 10^{34}$\\
26&20&&WN4b&14.69&--0.05&--0.09&0.17&--0.22&30&236&1.7$\times 10^{-12}$&0.20&0.21&9.9$\times 10^{35}$\\
31&25&&WN4b&15.39&0.03&--0.07&--0.17&--0.02&33&223&8.2$\times 10^{-13}$&0.16&0.33&7.5$\times 10^{35}$\\
35&27&&WN3(h)&14.84&--0.14&--0.18&0.01&--0.14&30&172&1.1$\times 10^{-12}$&0.02&0.12&4.8$\times 10^{35}$\\
37&30&&WN3o&16.43&0.21&0.07&--0.87&-&32&187&2.4$\times 10^{-13}$&-&0.58&35.2$\times 10^{35}$\\
40&33&&WN4(h)a&14.70&--0.20&--0.17&--0.08&--0.03&27&61&4.6$\times 10^{-13}$&0.12&0.11&1.9$\times 10^{35}$\\
41&35&&WN4b&14.88&--0.06&--0.05&0.05&--0.04&27&248&1.5$\times 10^{-12}$&0.01&0.20&8.7$\times 10^{35}$\\
46&38&&WN4o&15.27&--0.05&--0.05&--0.14&--0.10&28&131&5.4$\times 10^{-13}$&0.26&0.26&3.8$\times 10^{35}$\\
47&39&&WN3b&15.35&--0.10&--0.07&-&-&27&252&1.0$\times 10^{-12}$&-&0.15&4.8$\times 10^{35}$\\
48&40&&WN4b&14.69&--0.09&--0.07&--0.21&--0.06&32&240&1.8$\times 10^{-12}$&0.06&0.17&9.3$\times 10^{35}$\\
50&41&(b)&WN5h&14.49&--0.22&--0.18&-&-&21&32&2.9$\times 10^{-13}$&-&0.10&1.2$\times 10^{35}$\\
51&42&&WN3b&15.19&--0.15&--0.18&--0.25&0.09&46&265&1.2$\times 10^{-12}$&0.00&0.09&4.8$\times 10^{35}$\\
56&46&&WN4b&14.53&--0.10&--0.09&--0.15&-0.02&29&193&1.6$\times 10^{-12}$&0.05&0.23&1.0$\times 10^{36}$\\
57&45&&WN4b&14.87&--0.09&--0.08&--0.20&--0.14&29&262&1.6$\times 10^{-12}$&0.12&0.21&9.5$\times 10^{35}$\\
60&49&&WN4(h)a&14.29&--0.21&--0.23&-&-&30&39&4.2$\times 10^{-13}$&-&0.14&2.0$\times 10^{35}$\\
62&51&&WN3(h)&15.14&--0.18&--0.16&--0.15&--0.03&30&121&5.8$\times 10^{-13}$&0.04&0.17&3.0$\times 10^{35}$\\
63&52&&WN4ha:&14.58&--0.22&--0.22&--0.12&0.05&30&65&5.5$\times 10^{-13}$&0.09&0.11&2.3$\times 10^{35}$\\
65&55&&WN4o&15.51&0.14&0.08&-&-&29&162&4.8$\times 10^{-13}$&-&0.54&9.1$\times 10^{35}$\\
66&54&&WN3(h)&15.30&--0.19&--0.17&0.02&-&32&77&3.3$\times 10^{-13}$&-&0.12&1.4$\times 10^{35}$\\
67&56&&WN5ha&13.74&--0.02&--0.06&0.01&--0.06&27&40&6.6$\times 10^{-13}$&0.36&0.34&6.2$\times 10^{35}$\\
73&63&(b)&WN5ha&14.65&--0.16&--0.16&-&-&22&28&2.1$\times 10^{-13}$&-&0.18&1.1$\times 10^{35}$\\
74&63a&&WN3(h)a&15.50&--0.14&--0.14&-&-&36&45&1.5$\times 10^{-13}$&-&0.19&8.5$\times 10^{34}$\\
75&59&&WN4o&14.47&--0.11&--0.16&-&-&29&185&1.7$\times 10^{-12}$&-&0.17&8.8$\times 10^{35}$\\
81&65a&&WN5h&15.38&0.00&--0.13&-&-&19&37&1.4$\times 10^{-13}$&-&0.37&1.4$\times 10^{35}$\\
82&66&&WN3b&15.93&0.13&0.10&--0.37&0.16&32&243&5.0$\times 10^{-13}$&0.16&0.43&6.5$\times 10^{35}$\\
88&70a&(a)&WN4b/WCE&17.75&0.46&0.42&-&-&45&325&1.1$\times 10^{-13}$&-&0.81&5.3$\times 10^{35}$\\
94&85&(a)&WN4b&14.89&0.09&0.09&--0.12&-0.09&53&380&2.0$\times 10^{-12}$&0.25&0.35&2.0$\times 10^{36}$\\
117&88&(a)&WN5ha&12.95&--0.16&--0.13&--0.06&--0.06&33&42&1.5$\times 10^{-12}$&0.16&0.17&8.0$\times 10^{35}$\\
122&92&&WN5(h)&12.75&0.03&0.05&--0.25&--0.02&22&82&3.1$\times 10^{-12}$&0.43&0.44&4.2$\times 10^{36}$\\
128&96&&WN3b&15.46&0.00&--0.07&0.76&0.09&30&317&1.0$\times 10^{-12}$&-&0.26&7.2$\times 10^{35}$\\
131&98&&WN4b&14.27&--0.07&--0.14&&-&27&220&2.3$\times 10^{-12}$&-&0.20&1.3$\times 10^{36}$\\
132&99&&WN4b(h)&14.62&0.10&0.00&--0.29&0.03&31&301&2.1$\times 10^{-12}$&0.27&0.38&2.2$\times 10^{36}$\\
134&100&&WN4b&14.51&--0.05&--0.20&--0.10&--0.04&34&267&2.4$\times 10^{-12}$&0.11&0.11&1.0$\times 10^{36}$\\
\hline
WR&&&&&&&$\Delta b$ & &&&&MD01\\
\hline
1 & && WN3ha & 15.27 & --0.10 & --0.17 & 0.11 &  & 27 & 31.2 & 1.0$\times 10^{-13}$ & 0.28 & 0.26 & 1.1$\times 
10^{35}$ \\ 
2 & && WN5ha & 14.39 & --0.19 & --0.26 & 0.16 &  & 18 & 15.9 & 1.2$\times 10^{-13}$ & 0.17 & 0.12 & 8.3$\times 
10^{34}$ \\
4 & && WN6h  & 13.48 & --0.18 & --0.21 & 0.08 &  & 20 & 43.8 & 9.7$\times 10^{-13}$ & 0.16 & 0.16 & 7.8$\times 
10^{35}$ \\
9 &  && WN3ha& 15.39 & --0.25 & --0.26 &      &  & 28 & 23.2 & 9.1$\times 10^{-14}$ & 0.19 & 0.07 & 5.4$\times 
10^{34}$ \\
10&  &&WN3ha & 16.05 & --0.16 & --0.24 &      &  & 30 & 24.7 & 4.0$\times 10^{-14}$ & 0.24:& 0.17 & 3.3$\times 
10^{34}$ \\
11&  &&WN4ha+?& 15.17 & --0.12& 0.06   &      &  & 28 & 14.0 & 6.3$\times 10^{-14}$ & 0.50 & 0.47 & \\
  &  &&WN4ha & 15.70:& --0.14:&  --0.14:&     &  & 28 & 19.3 &                      &      & 0.20:& 5.8$\times 
10^{34}$ \\
12&&& WN3ha   & 16.17 & --0.21 & --0.21 &      &  & 28 & 22.1 & 4.1$\times 10^{-14}$ & 0.17 & 0.12 & 2.9$\times 
10^{34}$ \\
\hline
\end{tabular}
\begin{flushleft}
(a)~Observations taken at airmass $> 2$; 
(b)~Revised to WN5 from WN4 since N\,{\sc iv} $\lambda$4058 $\gg$ 
N\,{\sc v} $\lambda\lambda$4603--20/N\,{\sc iii} $\lambda\lambda$4634--41 
(Smith et al. 1996)
\end{flushleft}
\end{table*}

\clearpage

\begin{table*}
\begin{flushleft}
{\bf Table A2.} Photometric properties of WC stars in the LMC based upon
our Mt Stromlo 2.3m/DBS spectrophotometry, including C\,{\sc iv} 
$\lambda$5808 line measurements.
Interstellar extinction  corrections follow a standard Seaton (1979) 
reddening law. A distance of 49\,kpc and 62\,kpc is adopted for the 
LMC (see text). We compare synthetic magnitudes with 
 Torres-Dodgen \& Massey (1988, TM88) and reddenings with Smith et al. (1990,
S90). Spectral types are from Bartzakos et al. (2001).\\
\end{flushleft}
\begin{tabular}{c@{\hspace{2mm}}c@{\hspace{2mm}}
c@{\hspace{2mm}}l@{\hspace{1.5mm}}rrrrcrrrccc}
\hline
\multicolumn{2}{c}{Star}&&Sp&\multicolumn{1}{c}{\it v}
&\multicolumn{1}{c}{\it b-v}&\multicolumn{1}{c}{\it$\Delta$v}
&\multicolumn{1}{c}{\it$\Delta$(b-v)}&
FWHM&\multicolumn{1}{c}{W$_{\lambda}$}&$F_{5808}$&$E_{b-v}$&$E_{b-v}$ 
&$L_{5808}$\\
BAT99&Br&&Type&mag&mag&\multicolumn{2}{c}{-- TM88 --}& \AA& \AA& erg 
s$^{-1}$cm$^{-2}$ &S90&This work&erg\,s$^{-1}$\\
\hline
8&8& & WC4   & 14.87 & --0.05 & --0.02 & 0.08  & 54 & 1317 & 5.3$\times 10^{-12}$ & 0.19 & 0.12 & 2.2$\times 10^{36}$\\
9&7& & WC4   & 14.96 & --0.11 & --0.06 &--0.01 & 55 & 1658 & 5.4$\times 10^{-12}$ & 0.25 & 0.13 & 2.4$\times 10^{36}$\\
10&9&(a)& WC4+O9.5II&11.02&--0.28 &       &       & 62 & 74   &7.8$\times 
10^{-12}$  &      & 0.04 & 2.5$\times 10^{36}$\\
11&10& & WC4 & 13.87 & --0.08 & --0.02 & --0.02 & 73 & 1320 & 1.3$\times 10^{-11}$ & 0.20 & 0.09 & 4.8$\times 10^{36}$\\
20 & 16a&&WC4+O &14.01&0.01   &        &        & 75 & 264 & 1.9$\times 10^{-12}$ &      & 0.33 &  1.6$\times 10^{36}$\\
28 & 22 & & WC4+O5--6V-III & 12.05 & --0.26 & 0.03 & --0.07 & 67 & 183 & 7.7$\times 10^{12}$ & & 0.06 & 2.7$\times 10^{36}$\\
34 & 28 & & WC4+OB & 12.76 & --0.28 & --0.13 & --0.02 & 52 & 277 & 6.0$\times 10^{-12}$ &   & 0.04 & 1.9$\times 10^{36}$\\
38 & 31 & & WC4+O8I: & 11.37 & --0.26 &    &      & 70 & 96 & 7.5$\times 10^{-12}$ &    & 0.06 & 2.6$\times 10^{36}$ \\
39 & 32 & & WC4+O6V--III & 12.33 & --0.27 & --0.06 & --0.04 & 75 & 215 & 6.3$\times 10^{-12}$&& 0.05 & 2.1$\times 19^{36}$ \\
52 & 43 &  & WC4 & 14.03 & --0.11 &   &    & 61 & 1287 & 1.1$\times 10^{-11}$ & 0.07 & 0.09 & 4.4$\times 10^{36}$\\
53 & 44 & & WC4+OB & 13.04 & --0.19 &   &    & 72 & 525 & 8.3$\times 10^{-12}$ &  & 0.13 & 3.6$\times 10^{36}$\\
61 & 50 & & WC4 & 13.90 & --0.06 & --0.13 & 0.30 & 72 & 1218 & 1.4$\times 10^{-11}$ & 0.09 & 0.11 & 5.8$\times 10^{36}$\\
70 & 62 & & WC4+OB &13.93 & 0.11 & --0.07 & 0.25 & 83 & 880 & 7.3$\times 10^{-12}$ & 0.30 & 0.39 & 7.4$\times 10^{36}$\\ 
84 & 68 & & WC4+OB & 12.92 & --0.17 & & & 63 & 411 & 7.9$\times 10^{-12}$ &  & 0.15 & 3.7$\times 10^{36}$ \\
85 & 67 & & WC4+OB & 12.04 & 0.00 & & & 72 & 100 & 4.4$\times 10^{-12}$ & & 0.32 & 3.5$\times 10^{36}$ \\
90 & 74 & & WC4    & 15.42 & 0.12 & 0.01 & --0.07 & 57 & 1451 & 3.6$\times 10^{-12}$ & 0.39 & 0.28 & 2.6$\times 10^{36}$\\ 
121 & 90a & & WC4  & 17.21 & 0.31 & & & 59 & 1258 & 7.4$\times 10^{-13}$ &   & 0.59 & 1.4$\times 10^{36}$ \\
\hline
\end{tabular}
\begin{flushleft}
(a)~Walborn et al. (1999) have published spatially resolved blue optical
spectroscopy of the WC4 component (HD~32228-2) for which $b$=14.15 mag.
\end{flushleft}
\end{table*}

\end{appendix}

\end{document}